\documentclass[11pt,a4paper]{article}
\pdfoutput=1
\usepackage{jheppub}

\usepackage[T1]{fontenc}
\usepackage{graphicx}
\usepackage{comment}
\usepackage{grffile}

\usepackage{slashed}
\usepackage{amsmath,amsfonts,bbm,latexsym,amssymb}
\usepackage{epsfig}
\usepackage{array}

\usepackage{booktabs}
\usepackage[utf8]{inputenc}
\usepackage{fancyvrb}
\usepackage{subfig}
\usepackage{caption}
\usepackage{xcolor}

\usepackage{soul}

\usepackage{bm}
\usepackage{pifont}

\definecolor{darkgreen}{rgb}{0.13, 0.50, 0.09}

\newcommand{\newc}{\newcommand}
\newc{\be}{\begin{equation}}
\newc{\ee}{\end{equation}}
\newc{\bea}{\begin{eqnarray}}
\newc{\eea}{\end{eqnarray}}
\newc{\ol}{\overline}
\newc{\wt}{\widetilde}
\newc{\bs}{\boldsymbol}
\newc{\m}{\mathcal}
\newc{\la}{\langle}
\newc{\ra}{\rangle}

\newcommand{\beq}{\begin{eqnarray}}
\newcommand{\eeq}{\end{eqnarray}}
\newcommand{\bpmatrix}{\begin{pmatrix}}
\newcommand{\epmatrix}{\end{pmatrix}}

\renewcommand{\ol}{\text{1l}}

\newcommand{\figref}[1]{Fig.~\ref{#1}}
\renewcommand{\eqref}[1]{Eq.~(\ref{#1})}
\newcommand{\bib}[1]{Ref.~\cite{#1}}
\newcommand{\bibs}[1]{Refs.~\cite{#1}}
\newcommand{\tab}[1]{Table~\ref{#1}}
\newcommand{\sect}[1]{Section~\ref{#1}}

\newcommand{\bc}{\begin{center}}
\newcommand{\ec}{\end{center}}

\def\be{\begin{equation}}
\def\ee{\end{equation}}
\def\ba{\begin{eqnarray}}
\def\ea{\end{eqnarray}}
\def\ra{\rightarrow}

\def\m#1{m_{#1}}

\def\ltap{\;\centeron{\raise.35ex\hbox{$<$}}{\lower.65ex\hbox{$\simeq$}}\;}
\def\gtap{\;\centeron{\raise.35ex\hbox{$>$}}{\lower.65ex\hbox{$\simeq$}}\;}

\begin{document}
\begin{flushright}
CFTP/24-001
\end{flushright}
\vspace*{1cm}

\title{Impact of New Experimental Data on the C2HDM:
the Strong Interdependence between LHC Higgs Data
and the Electron EDM}

\author[a]{Thomas Biek\"{o}tter,}
\author[b]{Duarte Fontes,}
\author[a]{Margarete M\"{u}hlleitner,}
\author[c]{Jorge C. Rom\~ao,}
\author[d,e]{Rui Santos,}
\author[c]{and Jo\~{a}o P. Silva}

\affiliation[a]{Institute for Theoretical Physics,
Karlsruhe Institute of Technology, \\
76128 Karlsruhe, Germany}
\affiliation[b]{Department of Physics, Brookhaven National Laboratory,\\
Upton, N.Y., 11973, U.S.A.}
\affiliation[c]{Departamento de F\'{\i}sica and CFTP,
Instituto Superior T\'{e}cnico, Universidade de Lisboa, \\
Avenida Rovisco Pais 1, 1049-001 Lisboa, Portugal}
\affiliation[d]{ISEL - Instituto Superior de Engenharia de Lisboa, \\
Instituto Polit\'ecnico de Lisboa 1959-007 Lisboa, Portugal}
\affiliation[e]{Centro de F\'{\i}sica Te\'{o}rica e Computacional, Faculdade de Ci\^{e}ncias, \\
Universidade de Lisboa, Campo Grande, Edif\'{\i}cio C8,
1749-016 Lisboa, Portugal}
\emailAdd{thomas.biekoetter@kit.edu}
\emailAdd{dfontes@bnl.gov}
\emailAdd{margarete.muehlleitner@kit.edu}
\emailAdd{jorge.romao@tecnico.ulisboa.pt}
\emailAdd{rasantos@fc.ul.pt}
\emailAdd{jpsilva@cftp.ist.utl.pt}

\date{\today}

\abstract{
The complex two-Higgs doublet model (C2HDM) is one of the simplest
extensions of the Standard Model with a source of CP-violation in the
scalar sector.  It has a $\mathbb{Z}_2$ symmetry, softly broken by a
complex coefficient. There are four ways to implement this symmetry
in the fermion sector, leading to models known as Type-I, Type-II,
Lepton-Specific and Flipped. In the latter three models, there is
\textit{a priori} the surprising possibility that the
$125~\textrm{GeV}$ Higgs boson couples mostly as a scalar to top
quarks, while it couples mostly as a pseudoscalar to bottom quarks.
This ``maximal'' scenario was still possible with the data available
in 2017. Since then, there have been more data on the
$125~\textrm{GeV}$ Higgs boson, direct searches for CP-violation in
angular correlations of $\tau$-leptons produced in Higgs boson decays,
new results on the electron electric dipole moment, new constraints
from LHC searches for additional Higgs bosons and new
results on $b\to s \gamma$ transitions.
Highlighting the crucial importance of the physics results of LHC's
Run~2, we combine all these experiments and show that the ``maximal''
scenario is now excluded in all models. Still, one can have a
pseudoscalar component in $h \tau\bar{\tau}$ couplings in the
Lepton-Specific case
as large as $87\%$ of the scalar component for all mass orderings of the neutral scalar bosons.}

\maketitle

\section{Introduction}
\label{sec:intro}

The discovery of a $125~\textrm{GeV}$ neutral scalar ($h_{125}$) by
the ATLAS~\cite{Aad:2012tfa} and CMS~\cite{Chatrchyan:2012ufa}
collaborations at the Large Hadron Collider (LHC) opened a window into
the scalar sector. The subsequent Run~2 stage has greatly improved our
knowledge, as will become apparent in this article. However, there is
still much to uncover, as we start Run 3 and peer into the scalar
sector with unprecedented precision.

One interesting example concerns the CP properties of the $h_{125}$.  The
early detection of the $h_{125} \rightarrow Z Z^*$ decay meant that the
Higgs boson could not be a pure CP-odd
state~\cite{ATLAS-CONF-2013-013,CMS:2013fjq}.
But maybe it can be a mixture
of CP-even and CP-odd components.  More surprisingly, the $h_{125}$ can
couple as mostly CP-even to some states and as mostly CP-odd to
others.  Consider a two-Higgs doublet model with a $\Phi_1 \rightarrow
\Phi_1$, $\Phi_2 \rightarrow - \Phi_2$ symmetry, softly broken in
the potential by a term,
\be
V_\textrm{soft}
=
m_{12}^2 \Phi_1^\dagger \Phi_2 + m_{12}^{2\, *} \Phi_2^\dagger \Phi_1\, ,
\ee
where $m_{12}^2$ is a complex coefficient.  This is known as the
complex two-Higgs doublet model (C2HDM).  We will denote by ``real
2HDM'' the model obtained when all coefficients in the scalar
potential (including $m_{12}^2$) are real\footnote{In
\bib{Fontes:2021znm} it is argued that, since neutral meson
observables require explicit complex CP-violating dimension-four
Yukawa couplings to quarks, the real 2HDM might not be a fully
consistent theory.}; for reviews, see \bibs{Gunion:1989we,
Branco:2011iw}.  The real 2HDM was introduced by T.~D.~Lee in order to
show that one can have a spontaneous symmetry breaking origin for
CP-violation \cite{Lee:1973iz}.
In contrast, in the C2HDM investigated here the CP-violation appears
explicitly in the potential \cite{Ginzburg:2002wt,Weinberg:1990me}. 

Due to the presence of additional sources of CP-violation and the
possibility of accommodating a strong first-order electroweak phase
transition, the C2HDM is a suitable framework for an explanation of
the baryon asymmetry of the Universe by means of electroweak
baryogenesis.
This model has been studied extensively in the
literature; see, for example, \bibs{Khater:2003wq,
ElKaffas:2007rq, Grzadkowski:2009iz, Arhrib:2010ju, Barroso:2012wz,Abe:2013qla,
Inoue:2014nva, Cheung:2014oaa, Fontes:2014xva,Grober:2017gut,Fontes:2015xva, Fontes:2015mea,
Chen:2015gaa, Chen:2017com, Muhlleitner:2017dkd, Fontes:2017zfn, Basler:2017uxn,
Aoki:2018zgq, Basler:2019iuu, Wang:2019pet, Boto:2020wyf, Cheung:2020ugr,
Altmannshofer:2020shb, Fontes:2021iue, Basler:2021kgq, Frank:2021pkc, Abouabid:2021yvw,
Fontes:2022izp,Azevedo:2023zkg,Goncalves:2023svb}. 
In particular, a full analysis was performed of the C2HDM parameter
space consistent with the experimental data available at the end of
2017 \cite{Fontes:2017zfn}.  The 2017 analysis introduced a new code,
{\tt C2HDM\_HDECAY}, implementing the C2HDM in the well-known {\tt
HDECAY} program \cite{Djouadi:1997yw, Djouadi:2018xqq}, and used also:
\begin{itemize}
\item Signal strength constraints on $h_{125}$
from the combination of ATLAS and CMS data
collected at 7~TeV and 8~TeV~\cite{Khachatryan:2016vau};
\item HiggsBounds 4.3.1~\cite{Bechtle:2013wla}, for data from
  searches for additional scalars; 
\item The electron electric dipole moment (eEDM) limit of $8.7 \times
  10^{-29} \, \text{e.cm}$~\cite{ACME:2013pal}; 
\item The lower bound of 580 GeV on the charged Higgs boson mass,
  $m_{H^\pm}$, from radiative $B$-meson decays in the Type-II and
  Flipped models (introduced below) \cite{Misiak:2017bgg}. 
\end{itemize}

Since then, the experimental situation improved considerably on all four fronts. In
fact, there are new data on both the
properties of the $h_{125}$
(see \bib{ATLAS:2024fkg} for a recent summary of
the LHC Run~2 results from ATLAS) and
the searches for additional scalar states, a factor of roughly 20
improvement on the eEDM, and improved lower bounds on
$m_{H^\pm}$ in Type-II and Flipped.
In this paper, we analyze the impact
of the new experimental data on the parameter space
of the C2HDM. Specifically, we address the
question whether it is still experimentally
viable that the detected Higgs boson
at 125~GeV could be coupled to down-type
quarks and/or charged leptons as a dominantly
CP-odd state.\footnote{Similar analyses
focusing on the LHC Higgs data
have been carried out in the past
within an
effective field theory framework to describe
the Higgs-boson couplings,
see, e.g.~\bibs{Freitas:2012kw,Djouadi:2013qya,
Ferreira:2016jea,
Fuchs:2020uoc,Bahl:2022yrs}.}
To this end, we confront
the model with the following set of
recent measurements:
\begin{itemize}
\item The latest LHC data on the $h_{125}$ signal strengths,
including the full Run~2 data collected at~13~TeV, for the different
production and decay modes that have so far been detected. We
specifically use the ATLAS results summarized in Fig.~3 of
\bib{ATLAS:2022vkf}, demanding that the predicted signal rates agree
within $2\sigma$ with each individual signal-rate measurement.  The
ATLAS measurements are well in agreement with the corresponding CMS
results, such that all our conclusions would remain unchanged if
instead the CMS results or a combination of ATLAS+CMS results were
used;
\item The impact of the latest data of direct searches for
CP-violation by CMS 
using angular correlations in
decay planes of $\tau$ leptons produced in Higgs boson decays $
h_{125} \rightarrow \tau \bar{\tau}$ \cite{CMS:2021sdq}, setting
an upper limit of $\alpha_{h\tau\tau} < 41^\circ$ on the effective
mixing angle between the CP-even and CP-odd $\tau$-Yukawa coupling at
the $2\sigma$ confidence level (which, as we will show, has a very
strong impact on our analysis);\footnote{Our analysis
uses the CMS results, which was published
earlier than the corresponding ATLAS
results. ATLAS recently published a similar upper limit of
$\alpha_{h\tau\tau} < 34^\circ$~\cite{ATLAS:2022akr}.  Our conclusions
would remain unchanged if a combined CMS+ATLAS limit would be
considered.}
\item The impact of new searches for additional
scalars, as compiled in \texttt{HiggsBounds} 5.7.1 and 5.9.1
\cite{Bechtle:2008jh,Bechtle:2011sb,Bechtle:2013wla,Bechtle:2020pkv}
and in the newest \texttt{HiggsTools} 1.1.3 \cite{Bahl:2022igd},
incorporating the newest version~6 of \texttt{HiggsBounds}, extending
the previous versions by a large set of searches
that were performed including the full
Run~2 data collected at~13~TeV;
\item The recent 90\% confidence-level limit
on the eEDM of $1.1\times 10^{-29} \,
\text{e.cm}$
reported by the ACME collaboration~\cite{ACME:2018yjb}
and the 
most recent limit of $4.1\times
10^{-30} \, \text{e.cm}$
measured at JILA~\cite{Roussy:2022cmp};
\item
Updated bounds on the mass of the charged Higgs bosons
from measurements of radiative $B$-meson decays
(see the discussion in Sec.~\ref{subsec:Type-II}).

\end{itemize}
We note that in the C2HDM the
stringent eEDM bounds can only be evaded either
close to the CP-conserving limit of the model, or in scenarios where
cancellations between diagrams with different neutral scalar particles
occur~\cite{Jung:2013hka, Shu:2013uua,Hou:2023kho}.\footnote{The contributions
from the muon EDM and from non-leptonic EDMs
are currently less stringent 
\cite{Fontes:2017zfn} and will not be considered here.}
Also, henceforth HB stands for \texttt{HiggsBounds}
and HT for \texttt{HiggsTools}.

The paper is organised as follows.  In \sect{sec:datasets}, we
describe the parameter space of the model and the couplings of the
scalars to fermions and gauge bosons, and we present the theoretical
and experimental constraints used in our analysis.
In \sect{sec:CP-odd}, we discuss the current situation concerning the
possibility that the CP-odd components in the couplings of fermions to
$h_{125}$ are sizable compared to the respective CP-even components,
and thus potentially directly detectable at the LHC.  We summarize our
conclusions in \sect{sec:conclusions}.

\section{The C2HDM}
\label{sec:datasets}

\subsection{Physical parameters}
\label{subsec:parameters}

We follow closely the notation of \bib{Fontes:2017zfn}. In our notation, the vacuum expectation values (vevs) of the neutral
components of the scalar doublets are $ \langle \Phi_i^0 \rangle =
v_i/\sqrt{2}$ ($i=1,2$), where the parameters $v_i$ can be set to
be real and positive without loss of generality due to the freedom of
field re-definitions of the doublet fields $\Phi_i$, and $v^2 = v_1^2 + v_2^2 \simeq 246 \, \textrm{GeV}$, $\tan\beta= v_2/v_1$. The mixing of the neutral scalar particles can be described by three angles $\alpha_k$ ($k=1,2,3$), combined in the mixing matrix
\be
R =
\left(
\begin{array}{ccc}
c_1 c_2 & s_1 c_2 & s_2\\
-(c_1 s_2 s_3 + s_1 c_3) & c_1 c_3 - s_1 s_2 s_3  & c_2 s_3\\
- c_1 s_2 c_3 + s_1 s_3 & -(c_1 s_3 + s_1 s_2 c_3) & c_2 c_3
\end{array}
\right)\, ,
\label{matrixR}
\ee
where the short-hand notation $s_k \equiv \sin{\alpha_k}$ and $c_k
\equiv \cos{\alpha_k}$ has been used, and, without loss of generality,
\be
- \pi/2 < \alpha_1 \leq \pi/2,
\hspace{5ex}
- \pi/2 < \alpha_2 \leq \pi/2,
\hspace{5ex}
- \pi/2 \leq \alpha_3 \leq \pi/2.
\label{range_alpha}
\ee
We will make use of a mass-ordered notation in which the neutral
scalar masses obey $m_1 < m_2 < m_3$. Following \bib{Fontes:2017zfn}, we will describe the scalar sector of the C2HDM in terms of 9 independent parameters:
\begin{eqnarray}
v, \, \tan\beta, \, \alpha_1, \, \alpha_2, \, \alpha_3, \, m_{H^\pm}, \, m_1, \, m_2, \mbox{ and } \mbox{Re}(m_{12}^2) \;.
\end{eqnarray}
With this choice of independent
parameters, the mass of the heaviest neutral scalar, $m_3$, is a
dependent parameter, given by
\be
m_3^2 = \frac{m_1^2\, R_{13} (R_{12} \, t_\beta - R_{11})
+ m_2^2\ R_{23} (R_{22} \, t_\beta - R_{21})}{R_{33} (R_{31} - R_{32}
\, t_\beta)}\, , 
\label{m3_derived}
\ee
with $t_\beta \equiv \tan\beta$.
 Any of the three neutral scalars,
denoted $h_1$, $h_2$, $h_3$ in the
following, can in principle coincide with
$h_{125}$; we will thus explore the three possibilities:
$m_{h_{125}}=m_1$, $m_{h_{125}}=m_2$, and $m_{h_{125}}=m_3$.

The most general 2HDM suffers from potentially large flavour-changing
neutral scalar interactions with quarks, which could contribute to the
neutral meson mixing observables at levels much above what is
experimentally allowed. 
This can be
cured by the so-called natural flavour conservation mechanism, which
uses a $\mathbb{Z}_2$ symmetry, 
$\Phi_1 \rightarrow \Phi_1$, $\Phi_2 \rightarrow - \Phi_2$ extended to
the fermion sector in such a way that each of the three families of
fermions (up-type quarks, down-type quarks, charged leptons) couples
to one and only one scalar field
\cite{Glashow:1976nt,Paschos:1976ay}. Denoting by $\Phi_u$, 
$\Phi_d$, and $\Phi_\ell$ the doublet $\Phi_i$ ($i=1,2$) that couples to up-type quarks, down-type quarks, and charged leptons, respectively,
there are the following four
possibilities:\footnote{For all four possibilities, the Yukawa coupling
matrices of fermions with all neutral scalars are diagonal and
proportional to the fermion masses. Thus, besides the parameters for
the scalar sector, only fermion masses and the parameters of the
Cabibbo-Kobayashi-Maskawa (CKM) matrix (for the coupling to charged
scalars) are needed to specify a parameter point.}
\begin{itemize}
\item
Type-I:
$\Phi_u=\Phi_d=\Phi_\ell \equiv \Phi_2$\, ;
\item
Type-II:
$\Phi_u \equiv \Phi_2 \neq \Phi_d=\Phi_\ell \equiv \Phi_1$\, ;
\item
Lepton-Specific (LS):
$\Phi_u=\Phi_d \equiv \Phi_2 \neq \Phi_\ell \equiv \Phi_1$\, ;
\item
Flipped
$\Phi_u=\Phi_\ell \equiv \Phi_2 \neq \Phi_d \equiv \Phi_1$\, .
\end{itemize}
In the fermion and scalar mass bases,
the Yukawa Lagrangian for the neutral scalars
may be written as
\be
{\cal L}_Y = - \sum_{i=1}^3 \frac{m_f}{v} \bar{f}
\left[ c^e(h_i f \bar{f}) + i c^o(h_i f \bar{f}) \gamma_5 \right]\,
f\, h_i\, ,
\label{eq:yuklag}
\ee
where $f$ denotes the fermion field with mass $m_f$.
The real coefficients
$c^e(h_i f \bar{f})$ and $c^o (h_i f \bar{f})$ describe the CP-even
and CP-odd parts of the Yukawa couplings, respectively; we list them
in \tab{tab:1} in terms of the mixing matrix elements $R_{ij}$ (see
\eqref{matrixR}) and the mixing angle~$\beta$.
\begin{table}
\begin{center}
\begin{tabular}{rccc} \toprule
& $u$-type & $d$-type & leptons \\ \midrule
Type-I & $\frac{R_{i2}}{s_\beta} - i \frac{R_{i3}}{t_\beta} \gamma_5$
& $\frac{R_{i2}}{s_\beta} + i \frac{R_{i3}}{t_\beta} \gamma_5$ &
$\frac{R_{i2}}{s_\beta} + i \frac{R_{i3}}{t_\beta} \gamma_5$ \\
Type-II & $\frac{R_{i2}}{s_\beta} - i \frac{R_{i3}}{t_\beta} \gamma_5$
& $\frac{R_{i1}}{c_\beta} - i t_\beta R_{i3} \gamma_5$ &
$\frac{R_{i1}}{c_\beta} - i t_\beta R_{i3} \gamma_5$ \\
Lepton-Specific & $\frac{R_{i2}}{s_\beta} - i \frac{R_{i3}}{t_\beta} \gamma_5$
& $\frac{R_{i2}}{s_\beta} + i \frac{R_{i3}}{t_\beta} \gamma_5$ &
$\frac{R_{i1}}{c_\beta} - i t_\beta R_{i3} \gamma_5$ \\
Flipped & $\frac{R_{i2}}{s_\beta} - i \frac{R_{i3}}{t_\beta} \gamma_5$
& $\frac{R_{i1}}{c_\beta} - i t_\beta R_{i3} \gamma_5$ &
$\frac{R_{i2}}{s_\beta} + i \frac{R_{i3}}{t_\beta} \gamma_5$ \\ \bottomrule
\end{tabular}
\caption{Yukawa couplings of the Higgs
  bosons $h_i$ in the C2HDM, divided by the corresponding Standard
  Model Higgs couplings. The expressions correspond to 
  $[c^e(h_i f \bar{f}) +i c^o(h_i f \bar{f}) \gamma_5]$ from
  \eqref{eq:yuklag}. \label{tab:1}}
\end{center}
\end{table}
In the following, $c^e(h_{125} f \bar{f})$ and $c^o (h_{125} f
\bar{f})$ are abbreviated by $c^e_f$ and $c^o_f$, respectively.
Moreover, we represent the different families of up-type quarks,
down-type quarks and leptons by identifying $f$ with the generic
labels $t,b$ and $\tau$, respectively.  The effective mixing angle
between CP-even and CP-odd $\tau$-Yukawa couplings introduced in
Sec.~\ref{sec:intro} is then given by 
\begin{eqnarray}
\alpha_{h\tau\tau} =
\tan^{-1}|c_\tau^o| / |c_\tau^e| \;.
\end{eqnarray}

The LHC signal-rate measurements of~$h_{125}$ indicate that the
couplings of~$h_{125}$ to the massive gauge bosons $V=W,Z$ agree
within about 10\% with their Standard Model~(SM)
values~\cite{ATLAS:2022vkf,CMS:2022dwd}.  This favors the  
alignment limit of the~2HDM, in which the couplings of the neutral scalar $h_i$ identified with $h_{125}$, mimic the ones of the Higgs boson as predicted
by the~SM.  The deviations of the couplings of each neutral scalar
$h_i$ from that limit are parameterized by
\be
c(h_i VV) = c_\beta R_{i1} + s_\beta R_{i2}\, ,
\label{eq:c2dhmgaugecoup}
\ee
where in the exact alignment limit one finds $c(h_iVV) = 1$ for one of
 the three states $h_i = h_{125}$ and $c(h_iVV) =0$ for the other two $h_i\ne h_{125}$, and outside of the alignment limit $c(h_i VV)
< 1$ for all three neutral scalars.
We note that enforcing the alignment
limit in the ($Z_2$-symmetric)
C2HDM removes all
sources of CP-violation in the Higgs sector.
Thus, demanding the presence of CP-violation
and the alignment limit in the Higgs sector
requires an interpretation in a more general
2HDM, see \bib{Darvishi:2023nft} for a
recent discussion.
For reference, the full set of couplings in the C2HDM using our
notation is presented on the web page~\cite{C2HDM_FR}.

\subsection{Theoretical and Experimental Constraints}
\label{subsec:constraints}

Our input parameters are chosen as follows.  One of the neutral
scalars is identified with $h_{125}$; the masses of the remaining
neutral scalars are kept in the interval $30\mathrm{\; GeV}\leq
m_i<1\mathrm{\; TeV}$. As for $m_{H^\pm}$, we implement the bounds on
the $m_{H^\pm}-t_\beta$ plane arising from $B$ physics, most notably
those implied by the measurements of $B \to X_s \gamma$
\cite{Deschamps:2009rh,Mahmoudi:2009zx,Hermann:2012fc,Misiak:2015xwa,
  Misiak:2017bgg,Misiak:2020vlo}.  
We follow \bib{Fontes:2017zfn} in using $80 \mbox{ GeV} \le m_{H^\pm}
< 1 \mbox{ TeV}$ for the Type-I and LS models, and $580 \mbox{ GeV}
\le m_{H^\pm} < 1 \mbox{ TeV}$ for both Type-II and Flipped models
(although we will discuss the impact of the new bounds in the two
latter types).  As for the 
remaining input parameters, we also follow \bib{Fontes:2017zfn}, by
choosing the intervals: $0.8 \le t_{\beta} \le 35$, $- \frac{\pi}{2}
\le \alpha_{1,2,3} < \frac{\pi}{2}$ and $ 0 \le \mbox{Re}(m_{12}^2) <
500 000 \mbox{ GeV}^2$.
We then require our points to comply with the
measured values of the
oblique parameters $S$, $T$ and $U$~\cite{Branco:2011iw}
within $2\sigma$ of the experimental results quoted in
\bib{Baak:2014ora}, comparing against the
theoretically predicted values for the oblique
parameters at the one-loop level.
Finally, for each plot shown below, we will explicitly mention
which LHC constraints on $h_{125}$ are being used (whether those from
\bib{Khachatryan:2016vau} or those from \bib{ATLAS:2022vkf}),
which LHC constraints on extra scalars are being used (whether
HB-4.3.1 \cite{Bechtle:2013wla}, HB-5.9.1 \cite{Bechtle:2020pkv} or
HT-1.1.3 \cite{Bahl:2022igd}),
which eEDM constraints are being used (whether $8.7 \times 10^{-29} \,
\text{e.cm}$ \cite{ACME:2013pal,Hess:2014gxf}, $1.1\times 10^{-29} \,
\text{e.cm}$~\cite{ACME:2018yjb} or $4.1\times 10^{-30} \,
\text{e.cm}$~\cite{Roussy:2022cmp}),
and if the constraint on CP-violating
couplings coming from angular correlations
in the decays
$h_{125} \rightarrow \tau
\bar{\tau}$ \cite{CMS:2021sdq} is being used.
For the theoretical predictions of the eEDM
in the C2HDM, we follow \bib{Abe:2013qla}.

In all plots, we impose the known theoretical constraints, namely
boundedness from below and the non-existence of a lower lying
minimum~\cite{Ivanov:2015nea} to ensure the absolute stability of the
EW vacuum, and we demand perturbative unitarity~\cite{Kanemura:1993hm,
Akeroyd:2000wc,Ginzburg:2003fe}, applying an upper limit of $8 \pi$ on
the eigenvalues of the scalar four-point scattering matrix in the
high-energy limit.

\section{Searching for large CP-odd couplings}
\label{sec:CP-odd}

We perform an update of \bib{Fontes:2017zfn} of some of the authors of
this paper, looking in particular at the possibility that the $h_{125}
b \bar{b}$ and/or the $h_{125} \tau\bar{\tau}$ coupling might be
mostly CP-odd; that is, $|c^o_b| \gg |c^e_b|$ and/or $|c^o_{\tau}| \gg
|c^e_{\tau}|$.
In \bib{Fontes:2017zfn}, we found the situation 
summarized 
in \tab{tab:2}.
\begin{table}[h]
  \centering
  \begin{tabular}{|l|c|c|c|c|}\hline
    Type   & I & II& LS& Flipped\\ \hline
$h_1=h_{125}$&$\times$ &$\times$ &$\checkmark$ & $\checkmark$\\\hline
$h_2=h_{125}$ &$\times$ &$\checkmark$  &$\checkmark$ & $\times$\\\hline
$h_3=h_{125}$ &$\times$ &$\times$ &$\checkmark$ & $\times$\\\hline
  \end{tabular}
  \caption{Results for the
    possibility of sizable CP-odd components in the couplings of the
    Higgs boson at~125 GeV from \bib{Fontes:2017zfn}.  A checkmark
    (cross) means that it was (not) possible to have large CP-odd
    components $|c_f^o| > |c_f^e|$ ($f=b,\tau$) in the couplings of
    $h_{125}$.}
  \label{tab:2}
\end{table}

The impossibility to accommodate
mostly CP-odd couplings of $h_{125}$
found in all \mbox{Type-I} cases is easy to understand. As can be seen
in the first row of \tab{tab:1}, the CP-odd coupling
components in that Type are
always proportional to $R_{i3} /t_{\beta}$. 
Now, on the one hand, the $B$ physics constraints force $t_{\beta} >
1$. On the other hand, $R_{i3}$ is just a product of sine and cosine
of the rotation angles matrix of \eqref{matrixR}, so that
$|R_{i3}|\leq1$. More than this,
$|R_{i3}|$ is
further 
constrained by
$\mu_{VV}$ (the ratio between the new physics and the SM
value of the product between Higgs-boson
production and its decay to vector
bosons); the reason is that $|R_{i3}|$ 
is a measure of the CP-odd 
admixture of the state $h_i$,
thus suppressing the couplings to gauge bosons~\cite{Fontes:2015mea}.
Finally, since in Type-I all fermions couple as
the top quark to the scalars, this Type is
precluded from large CP-odd
components.\footnote{There are also direct measurements on the CP-odd
  \textit{vs.}~CP-even components of the top coupling coming from $pp
  \to t \bar{t} (h_{125} \to \gamma \gamma)$~\cite{ATLAS:2020ior},
  giving rise to $\alpha_{htt} < 43^\circ$
at 95\% confidence level, thus directly excluding
  the maximally CP-odd scenario with
  $|c_t^o| \gg |c_t^e|$.
  We also point out that the valid cases shown in \tab{tab:2} also
allowed for the so-called wrong-sign regime,
in which the Yukawa coupling is real but has the opposite sign it
would have in the SM 
\cite{Carmi:2012yp, Chiang:2013ixa, Ferreira:2014naa, Fontes:2014xva,
  Fontes:2014tga, Ferreira:2014dya}.}
As a consequence, in the following discussion
we do not consider the Type-I anymore and focus
on the other three Yukawa types of the C2HDM.

In the following subsections, we present several figures. In all of
them, the light green points are consistent with the old eEDM
of $8.7 \times 10^{-29} \, \text{e.cm}$, the dark green points with
the more recent result $1.1 \times 10^{-29} \, \text{e.cm}$ and the
dark red points with the new result, $4.1\times 10^{-30} \,
\text{e.cm}$. 
Also, the signs of $c^{e}_f$ and $c^{o}_f$ have no absolute meaning;
they are relative to the sign of  $k_V \equiv c(h_{125}VV)$, 
which is thus also taken into account in the plots by always plotting
$\text{sgn}(k_V) c_f^o$ \textit{vs.}~$\text{sgn}(k_V) c_f^e$.

\subsection{\label{subsec:Type-II}Type-II}

Reference~\cite{Fontes:2017zfn} found that, in Type-II, it was possible to
have sizable CP-odd components in the Yukawa couplings of $h_{125}$
for the case of $h_2=h_{125}$ (and only in this case).  In the
following figures, we reproduce this result, and investigate the
impact of recent LHC data.  We thus assume $h_2=h_{125}$, and we use
the  bound of $m_{H^\pm} > 580$
GeV~\cite{Misiak:2017bgg} resulting from measurements of $b \to s
\gamma$ transitions, which was the limit applied in
\bib{Fontes:2017zfn}. 
Meanwhile, there was an updated NNLO calculation
of the inclusive $b \to s \gamma$ branching ratios,
giving rise to a limit of $m_{H^\pm} > 800$~GeV~\cite{Misiak:2020vlo}. 
This limit was based on the HFLAV average value
$\mathrm{BR}(b\to s \gamma) = (3.32 \pm 0.15) \times
10^{-4}$ from 2019~\cite{HFLAV:2019otj}.
The current average value 
from HFLAV 2022~\cite{HFLAV:2022esi},
$\mathrm{BR}(b\to s \gamma) = (3.49 \pm 0.19)
\times 10^{-4}$, is slightly larger and has
a larger uncertainty.\footnote{Not including
the most recent Belle-II
measurement~\cite{Belle-II:2022hys}, the HFLAV
2022 average value is in good agreement with this
latest experimental result.}
A reanalysis of the impact of this new
result by the group that provided the
previous limits~\cite{Misiak:2017bgg,Misiak:2020vlo}
lowers the limit on the charged
Higgs-boson mass to $m_{H^\pm} > 500$~GeV at
$2\sigma$ confidence
level~\cite{Steinhauser:2023}.
Since the underlying experimental data
has undergone significant changes in recent years,
leading to substantial fluctuations in the
lower limit on the mass of the charged Higgs bosons,
we decided to keep the somewhat stronger limit of
$m_{H^\pm} > 580$ GeV~\cite{Misiak:2017bgg} applied
in the previous analysis~\cite{Fontes:2017zfn}
to facilitate a better comparison and a more direct
analysis of the impact of the other experimental
constraints applied in our analysis.
We have verified that 
our conclusions regarding the possibility of
accommodating sizable CP-odd components
in the couplings of~$h_{125}$ in the Type-II
and the Flipped Type do not depend on
whether a lower limit of 500~GeV or 580~GeV
is applied.

\begin{figure}[htb!]
  \centering
  \begin{tabular}{cc}
    \includegraphics[width=0.45\textwidth]{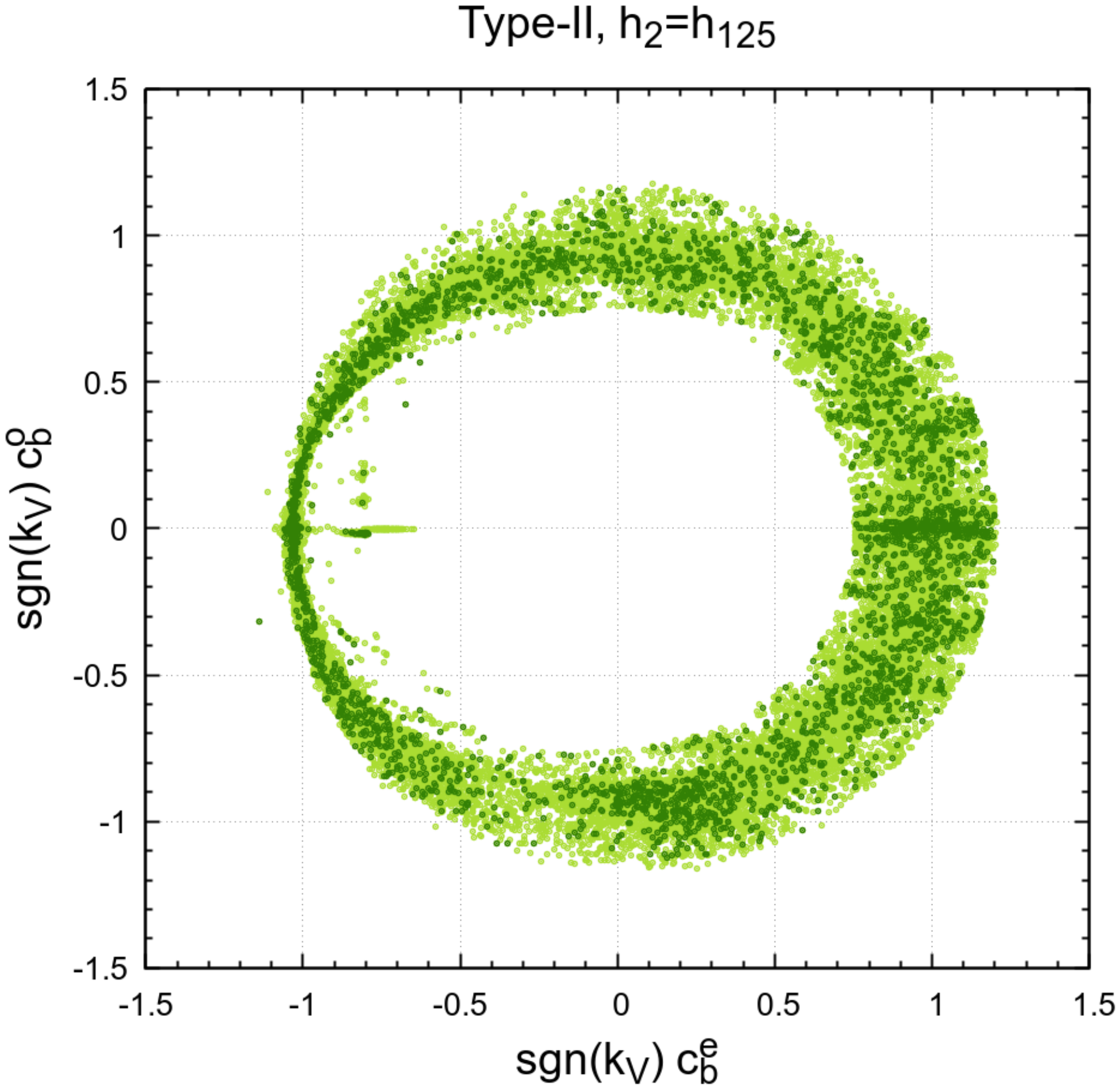}&
    \includegraphics[width=0.45\textwidth]{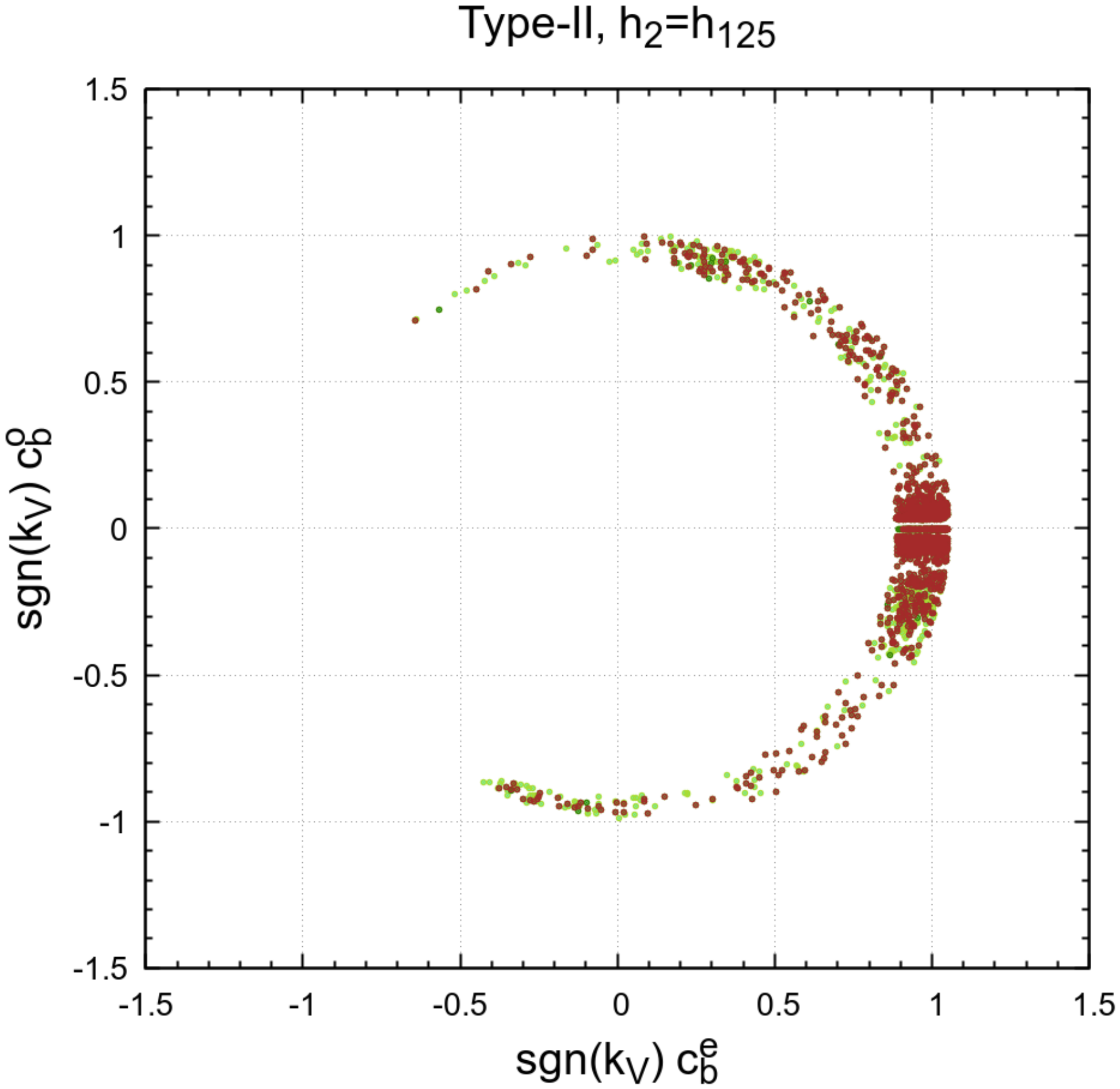}
  \end{tabular}
  \caption{CP-odd \textit{vs.}~CP-even component in the
    $h_{125}b\bar{b}$ coupling of allowed
    parameter points in Type-II, assuming
    $h_2=h_{125}$. Left panel: LHC 2017 data
    on $h_{125}$ and constraints from beyond-SM (BSM)
    scalar searches at 7~and 8~TeV
    using HB-4.3.1. Right
    panel: LHC 2022 data on $h_{125}$
    and constraints from BSM scalars
    including 13~TeV data using HT-1.1.3.
    The light green points are consistent
    with the old eEDM of
    $8.7 \times 10^{-29} \,
    \text{e.cm}$~\cite{ACME:2013pal,Hess:2014gxf},
    the dark green points with the more recent ACME
    result $1.1 \times 10^{-29}
    \, \text{e.cm}$~\cite{ACME:2018yjb}.
    The dark red points obey
    the currently strongest limit on the
    eEDM $4.1\times 10^{-30} \,
    \text{e.cm}$ reported by JILA~\cite{Roussy:2022cmp}.
    The fermion masses in the loops
    of diagrams contributing to the eEDM
    were taken as pole masses.
    The limit $\alpha_{h\tau\tau} <
    41^\circ$~\cite{CMS:2021sdq} from
    searches for CP-violation in angular
    correlations of $\tau$ leptons in $h_{125} \to \tau
    \bar \tau$ decays has not been applied in either
    of the plots in this figure.}
  \label{fig:1}
\end{figure}

Figure~\ref{fig:1} shows 
the allowed parameter space in the plane CP-odd ($c_b^o$) \textit{vs.}~CP-even ($c_b^e$)
 component of the $h_{125}b\bar{b}$
coupling. The left panel considers the
7~and 8~TeV LHC data for $h_{125}$
collected until 2017~\cite{Khachatryan:2016vau}
and the cross-section
limits from BSM scalar searches as implemented in
HB-4.3.1~\cite{Bechtle:2013wla},
which are the limits that were applied
in the previous analysis~\cite{Fontes:2017zfn}.
The right 
one considers additionally the 13~TeV LHC
data for $h_{125}$ collected during
Run~2~\cite{ATLAS:2022vkf}
and the cross-section limits
from BSM scalar searches at 13~TeV
as implemented in HT-1.1.3~\cite{Bahl:2022igd}.
The light green points are in agreement
with the old eEDM of
$8.7 \times 10^{-29} \, \text{e.cm}$~\cite{ACME:2013pal,Hess:2014gxf},
the dark green points with the more recent result $1.1 \times 10^{-29}
\, \text{e.cm}$~\cite{ACME:2018yjb}.
The dark red points obey
the current limit on the
eEDM~\cite{Roussy:2022cmp}.
The considerable reduction of the allowed parameter
space in the transition from the left to the right
plot is mainly caused by
the 
13~TeV data on $h_{125}$ collected
until 2022 (and 
to a lesser extend by improved searches for
heavier scalars involved in the different versions of HB). 
The width of the rings is related to the decay width
$\Gamma(h_{125}\to b \bar{b}) \propto (c_b^e)^2+ (c_b^o)^2$,
which is the largest contribution to the total
width and can thus also be responsible for modifications
in other decay channels, and the
new data reduces the allowed range for
$(c_b^e)^2+ (c_b^o)^2$.
Note also that the points close to the wrong sign limit
in the left panel of
\figref{fig:1} disappear with the new data,
which is mainly related to more precise
measurements in the gluon-fusion production
channels of~$h_{125}$.\footnote{See
Refs.~\cite{Su:2019ibd,Biekotter:2022ckj,
ATLAS:2024lyh}
for more detailed discussions about the
wrong sign limit in view of the
13~TeV LHC Higgs data.}
Yet, even after applying all current
constraints except for the direct limit
on the effective mixing angle $\alpha_{h\tau\tau}$
(see discussion below), but including the new limit on the
eEDM~\cite{Roussy:2022cmp}, we find allowed parameter
points in the Type-II with a large pseudoscalar
component in the
couplings of~$h_{125}$ to bottom quarks.

So far, we have not yet applied
the recent direct bound on a CP-odd coupling component from the
angular correlations of $\tau$ leptons in $h_{125} \rightarrow \tau
\bar{\tau}$ decays~\cite{CMS:2021sdq,ATLAS:2022akr}.
In the Type-II model, the down-type quarks are coupled to the
neutral scalars in the same way as the charged leptons,
such that $c_b^{e,o} = c_\tau^{e,o}$.
It follows
that, in this type, the
recent bound $ \alpha_{h\tau\tau} < 41^\circ$
has to be taken into account in
the study of the CP properties of
$h_{125}b\bar{b}$.
The limit on $\alpha_{h\tau\tau}$ has not been
applied in either of the plots in \figref{fig:1}.
Requiring that $\alpha_{hbb} < 41^\circ$,
with $\alpha_{hbb}= \tan^{-1} |c^o_b| / |c^e_b| = \alpha_{h\tau\tau}$,
excludes the possibility of $|c^o_b| \gg |c^e_b|$
in the right panel of 
\figref{fig:1}.
Nevertheless, 
the interesting possibility that $|c^o_b| \simeq |c^e_b|$
(and therefore also $|c^o_\tau| \simeq |c^e_\tau|$)
would still be allowed.

The above conclusions in the Type-II crucially
depend on a significant fine-tuning of the
model parameters in order to be compatible
with the stringent experimental upper bounds
on the eEDM. 
These limits can be evaded only as a result of
a cancellation between different contributions
to the eEDM at two-loop
level in the perturbative expansion
(as discussed in more detail below).
This cancellation gives rise to a strong
dependence of the predicted eEDM on the
model parameters, including the values for
the masses of the fermions that appear
as virtual particles in the loops
of Barr-Zee type diagrams~\cite{Barr:1990vd}.
The corresponding amplitudes
are proportional to the mass of the fermion
appearing in the loop. Consequently, the
numerically relevant contributions stem from
diagrams with an internal top quark, bottom
quark, or $\tau$ lepton.
At the two-loop level, it is formally consistent
to choose different renomalization prescriptions
for the fermion masses~\cite{Athron:2021evk},
and different approaches have been applied
in the literature.
The two most common choices have
been to use either $\overline{\rm MS}$
running masses
at the scale~$M_Z$
($\overline{m}_t(M_Z),
\overline{m}_b(M_Z),
\overline{m}_\tau(M_Z)$),
see e.g.~\bibs{Abe:2013qla,
King:2015oxa,Altmannshofer:2020shb},
or pole masses
for top quark and $\tau$ lepton in combination
with the running bottom-quark mass
at the scale $\overline{m}_b$
($m_t,\overline{m}_b(\overline{m}_b),
m_\tau$), see e.g.~\bibs{Fontes:2017zfn,
Basler:2017uxn,
Basler:2021kgq,
Bahl:2022yrs,Goncalves:2023svb}.
In the analysis discussed above, we have used
the latter possibility for the eEDM
predictions. In the following, we will
discuss the modifications resulting from
choosing the running masses at the scale $M_Z$.

\begin{figure}[htb]
  \centering
  \begin{tabular}{cc}
\includegraphics[width=0.45\textwidth]{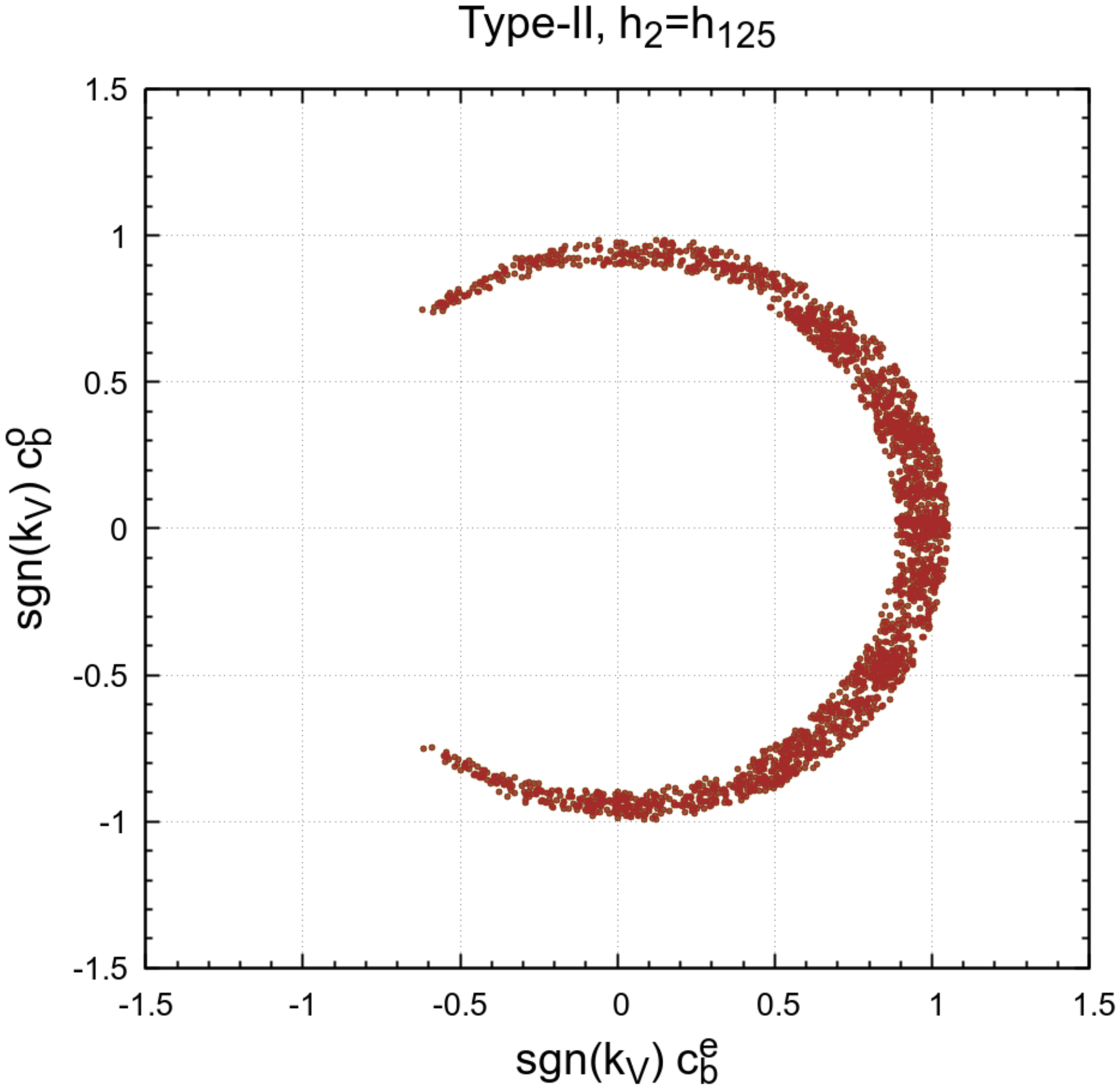}&
    \includegraphics[width=0.45\textwidth]{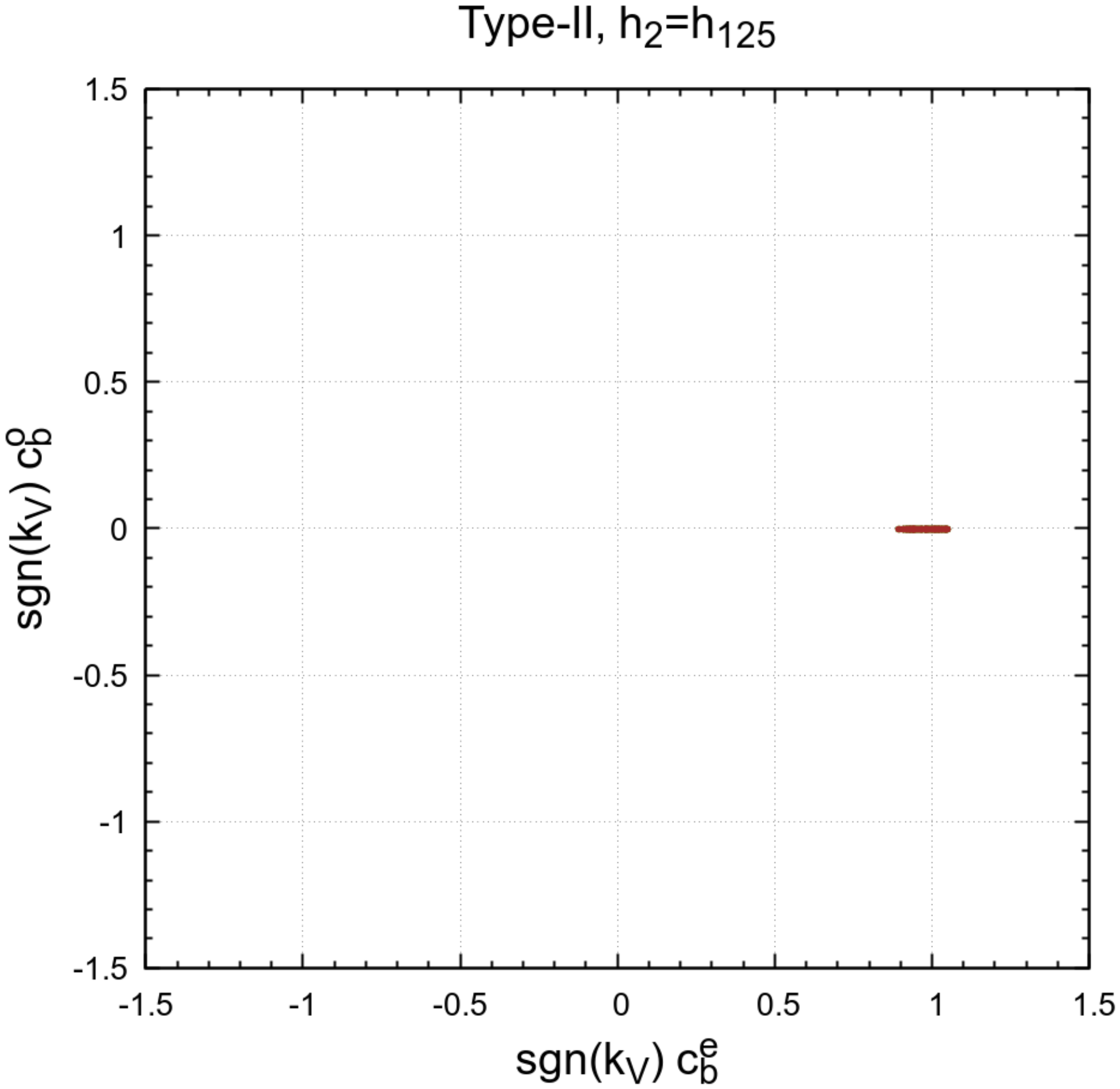} 
  \end{tabular}
  \caption{
  CP-odd \textit{vs.}~CP-even component in the $h_{125}b\bar{b}$
  coupling of allowed parameter points
  in Type-II, assuming $h_2=h_{125}$.
  All points obey the current experimental limit on the
  eEDM~\cite{Roussy:2022cmp}, where here
  the masses of the fermions
  in the loops of diagrams contributing to
  the eEDM were taken to be the running masses
  at the $M_Z$ scale (see text for details).
  Also applied are the constraints from
  the $h_{125}$ cross section measurements
  using LHC 2022 data collected at 13~TeV.
  The left panel does not include the LHC
  constraints on the extra
  scalars while in the
  right panel these constraints are applied
  including the most recent searches at 13~TeV
  using HT-1.1.3.} 
  \label{fig:2} 
\end{figure}

To this end, we generated a new set of
parameter points in the Type-II which all
satisfy the current experimental limit on
the eEDM with the eEDM
computed using the running masses at~$M_Z$.
Moreover, the parameter points fulfill the
other experimental and theoretical constraints
discussed in Sec.~\ref{subsec:constraints},
with the exception of the constraints from
BSM scalar searches at the LHC and from direct searches for CP-violation
in $h_{125} \to \tau \bar \tau$ decays.
The resulting parameter points are shown
in the plane of CP-odd {\it vs.}~CP-even
components of the $h_{125} b \bar b$ coupling
in the left plot of \figref{fig:2}.
One can see that, before the cross-section
limits from the LHC searches are applied,
the results are very similar to the case
shown in the right plot of \figref{fig:1},
where the eEDM was computed using the on-shell (OS)
prescription for the 
top-quark and $\tau$-lepton 
masses in combination with
$\overline{m}_b(\overline{m}_b)$.
However, after applying the LHC constraints
from searches for additional scalars
(see the discussion below for details), the only
still viable parameter points are situated
very close to the alignment limit, as is
shown in the right plot of \figref{fig:2}.
Hence,
if the eEDM is computed
using the running masses at the scale~$M_Z$,
we find that it is incompatible to
have both sizable CP-odd components in the
$h_{125} b \bar b$ coupling and agreement with the experimental upper limit
on the eEDM and with cross-section limits
from BSM scalar searches.
This is in clear contrast to our observations
using pole masses
$m_t$ and $m_\tau$ in combination with
$\overline{m}_b(\overline{m}_b)$
for the eEDM predictions,
as becomes apparent by comparing
the right plot of \figref{fig:2} to the right plot
of \figref{fig:1}.

\begin{figure}[htb]
  \centering
  \begin{tabular}{cc}
\includegraphics[width=0.45\textwidth]{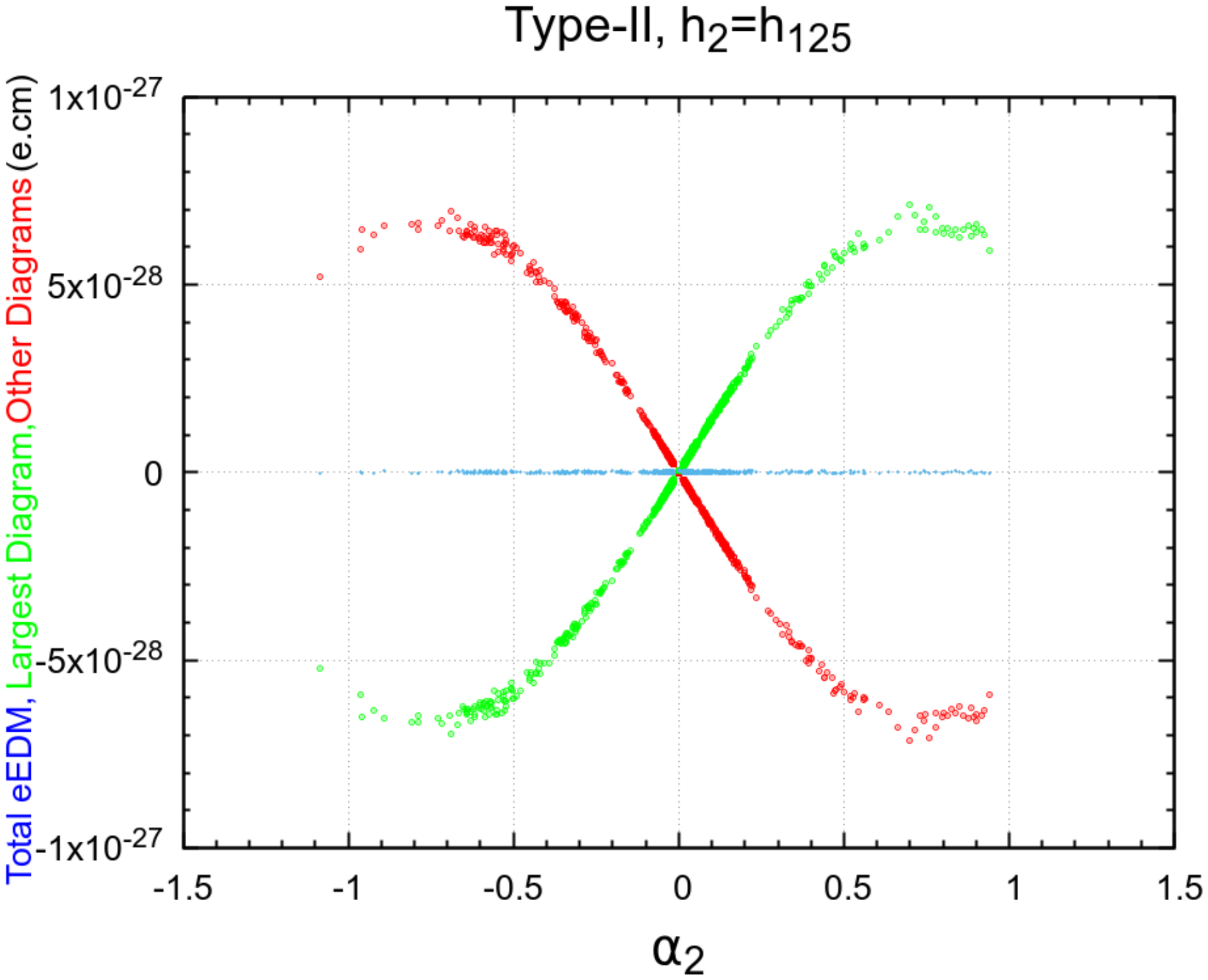}&
    \includegraphics[width=0.45\textwidth]{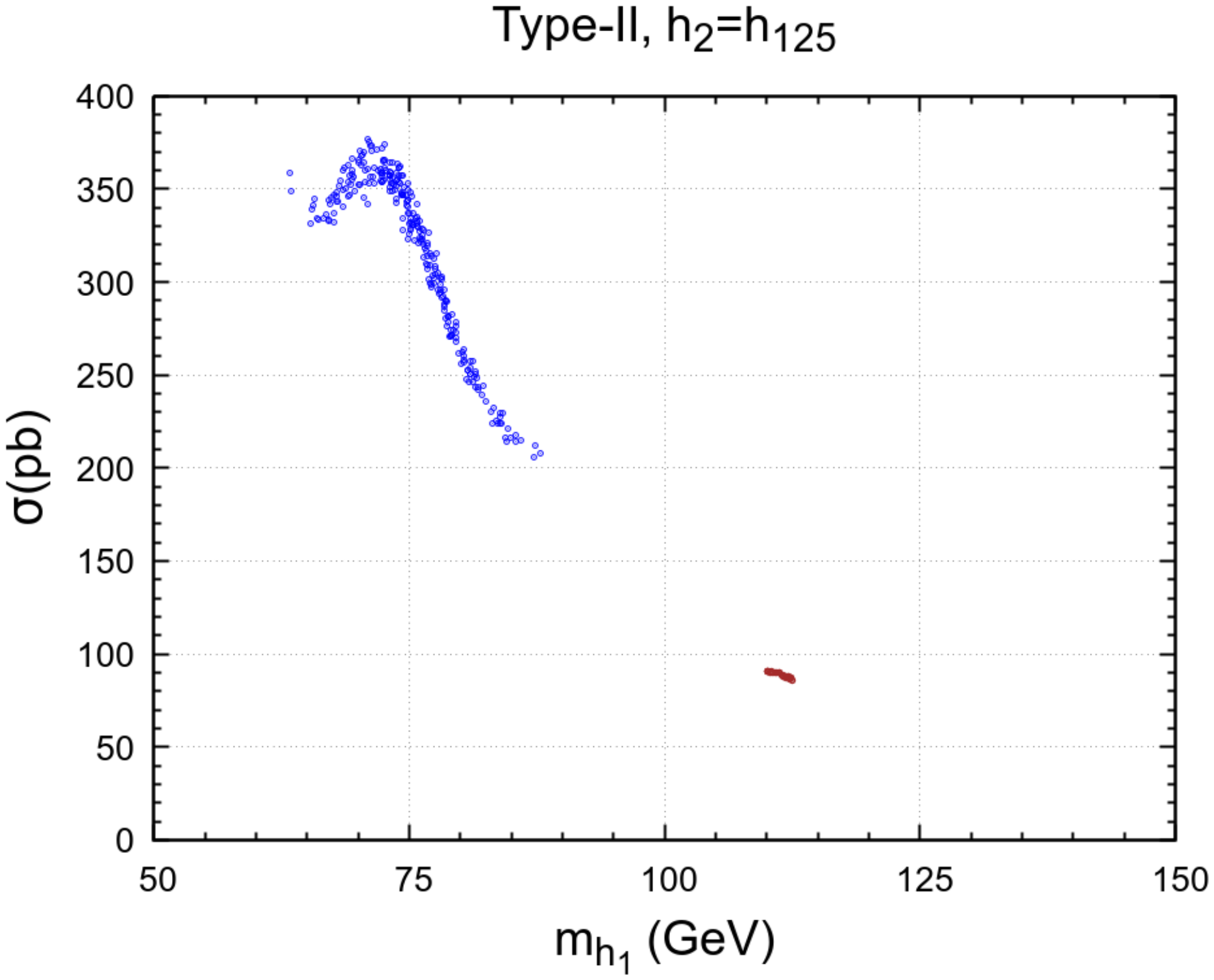} 
  \end{tabular}
  \caption{Left panel: Individual contributions to the eEDM
  taken into our account in our analysis.
  The green points show the contribution from one of the diagrams
  in the $W$-loop class, which gives the largest contribution to
  the eEDM, and the red points show the
  sum of 
  the remaining contributions.
  The blue points show the total eEDM
  (see text for the details).
    Right panel: LHC 13~TeV cross section
  for the gluon-fusion production of $h_1$
    as a function of
  $m_{h_{1}}$.
    Blue and brown points
  are the subset of
  the points shown in the right plot
  of \figref{fig:1} and in the left plot of
  \figref{fig:2}, respectively, that additionally
  satisfy the condition $|c_b^e| < 0.1$. 
      } 
  \label{fig:2new} 
\end{figure}

To gain more insight as to why the conclusions
in the Type-II model
depend so strongly on the choice of the
parameters (such as the precise values for
the fermion masses), we investigated the individual
Barr-Zee contributions to the eEDM.
Following the nomenclature of \bib{Abe:2013qla}, these
can be divided into four classes denoted
``fermion loops'', ``charged Higgs loops'',
``$W$ loops'' and ``$H^\mp W^\pm \gamma$ loops''.
In our scans, we observed that, for parameter points
that satisfy the experimental upper limit of
$4.1 \cdot 10^{-30}$~e.cm, the contributions from
the individual pieces can still be of the order
of $10^{-28}$~e.cm. This is illustrated in the left plot
of \figref{fig:2new}, where we show in green
the contribution from one of the diagrams in the $W$-loop class,
which typically is the numerically most
important piece,
and in red the sum of the other three contributions.
Finally, we show in blue the total eEDM containing
all four pieces, confirming the very significant fine-tuning
between the different components.

In our scans, this fine-tuning is achieved by
carefully adjusting the parameters of the model.
If one changes the fermion masses in the loops (e.g.~as a result of choosing different renormalization
prescriptions as discussed above), one has to adjust
the other parameters of the model in order
to maintain the fine-tuning in the eEDM prediction.
This concerns mainly the masses of the BSM scalars,
which then turns out to have a large impact on
the LHC phenomenology and corresponding constraints
on the C2HDM.
The difference of results in Type-II
according to the different prescription of the
fermion masses, as discussed above, has its origin
in the application of cross-section limits from
LHC searches --- more specifically, searches for scalar resonances produced via gluon-fusion with subsequent decay in
$\tau \bar \tau$ final states performed by CMS
at 13~TeV~\cite{CMS:2022goy}.\footnote{The corresponding
ATLAS search~\cite{ATLAS:2020zms}
does not include the mass region
below 125~GeV that is relevant in this discussion.}
In the right plot of \figref{fig:2new}, we show the
parameter points from the two scans that we
have performed in Type-II, with the
lightest scalar mass $m_{h_1}$ on the horizontal axis
and the gluon-fusion production cross section
for $h_1$ at 13~TeV on the vertical axis. Note that we only show the points that satisfy the additional condition
$|c_b^e| < 0.1$, which ensures that, for
the shown parameter points, the $h_{125} b \bar b$
coupling is mainly CP-odd.
The brown points are from the scan in which we used
the pole masses 
$m_t$ and $m_\tau$ and the $\overline{\rm MS}$ mass
$\overline{m}_b(\overline{m}_b)$
in the computation
of the eEDM (corresponding to a subset of the points
depicted in the right plot
of \figref{fig:1}), and the blue points  are
from the scan in which the running fermion masses
at the weak scale were used (corresponding to a subset of
the points depicted in the left plot
of \figref{fig:2}).
One can see that, for the latter set of parameter
points, the masses of the lightest scalar $h_1$ are lower
compared to the points for which the pole masses
$m_t$ and $m_\tau$ in combination with
$\overline{m}_b(\overline{m}_b)$
were used for the eEDM.
As a consequence, $h_1$ has
substantially larger production cross sections
at the LHC, and would have been observed
in searches for low-mass scalar resonances decaying
into $\tau$-lepton pairs.\footnote{The branching ratios
of the decay $h_1 \to \tau \bar \tau$ have values in the interval
8.6\% to 9.7\% for the brown points and 8.6\% to 9.2\%
for the blue points.
Therefore, the exclusion comes mainly from
the differences in the production cross section. 
}
The application of the corresponding cross-section
limits gives rise not only to the difference between the
left and the right plot of \figref{fig:2}, but also to
the different conclusions (regarding the possibility
of detectable CP-odd components in the couplings
of $h_{125}$) according to the mass renormalization
of the virtual fermions in diagrams contributing to the eEDM.

\subsection{\label{subsec:LS}Lepton-Specific}

In the LS model, the down-type quarks are coupled to the scalars
in the same way as the up-type quarks. As a consequence, 
given that (as discussed above) the $h_{125} t \bar{t}$
coupling is measured to be mainly 
CP-even, one can only 
find sizable CP-odd components in the LS model
in the coupling to leptons, i.e. $|c^o_{\tau}| \gg
|c^e_{\tau}|$.  In the following, we consider the three 
possible mass hierarchies with 
the lightest, the second-lightest or the heaviest
neutral scalar playing the role of~$h_{125}$.
With the 7 and 8~TeV LHC Higgs data collected
until 2017 and for the LS case, all placements
of~$h_{125}$ in the neutral
scalar mass orderings were still consistent with a
mostly CP-odd $h_{125} \tau\bar{\tau}$ 
coupling~\cite{Fontes:2017zfn}.

In view of the discussion in
Sec.~\ref{subsec:Type-II}, we computed the eEDM in the LS model using the running fermion masses at the $M_Z$ scale. However, we note that, in contrast to the Type-II model, the LS model is such that the fine-tuning of parameters
required to satisfy the upper bounds on the eEDM is not very severe. It follows that our findings are unchanged if,
instead of the running fermion masses at the $M_Z$ scale, we were to use the pole masses for the top quark and
the $\tau$ lepton as well as $\overline{m}_b(\overline{m}_b)$ (we verified this explicitly with dedicated scans).
The LS model has a further advantage over
the Type-II model, which concerns the fact that constraints from
collider experiments are less severe.
As discussed in Sec.~\ref{subsec:constraints}, in fact, the lower limit $m_{H^\pm} > 580$~GeV on the charged Higgs-boson mass
from measurements of $b \to s \gamma$ transitions
applicable in Type-II is not valid for
the LS model. Moreover, the LS model can realize
CP-violating effects only in the
couplings of the 125~GeV Higgs boson
to leptons, whereas its couplings to
quarks can remain at the same time approximately
CP-conserving. As a result, the LS model allows
for a sizable amount of CP-violation in the
couplings of the $h_{125}$ to leptons, without modifications
from CP-violating effects
to the most important production and decay channels
at the LHC. Thus, there is more freedom in this
type for the presence of CP-violating couplings
in regards to the cross-section
measurements of the detected Higgs boson.

In the following, we will analyze separately
the three possible mass hierarchies in the
LS model.

\subsubsection{$h_1=h_{125}$}

\begin{figure}[htb]
  \centering
  \begin{tabular}{cc}
    \includegraphics[width=0.45\textwidth]{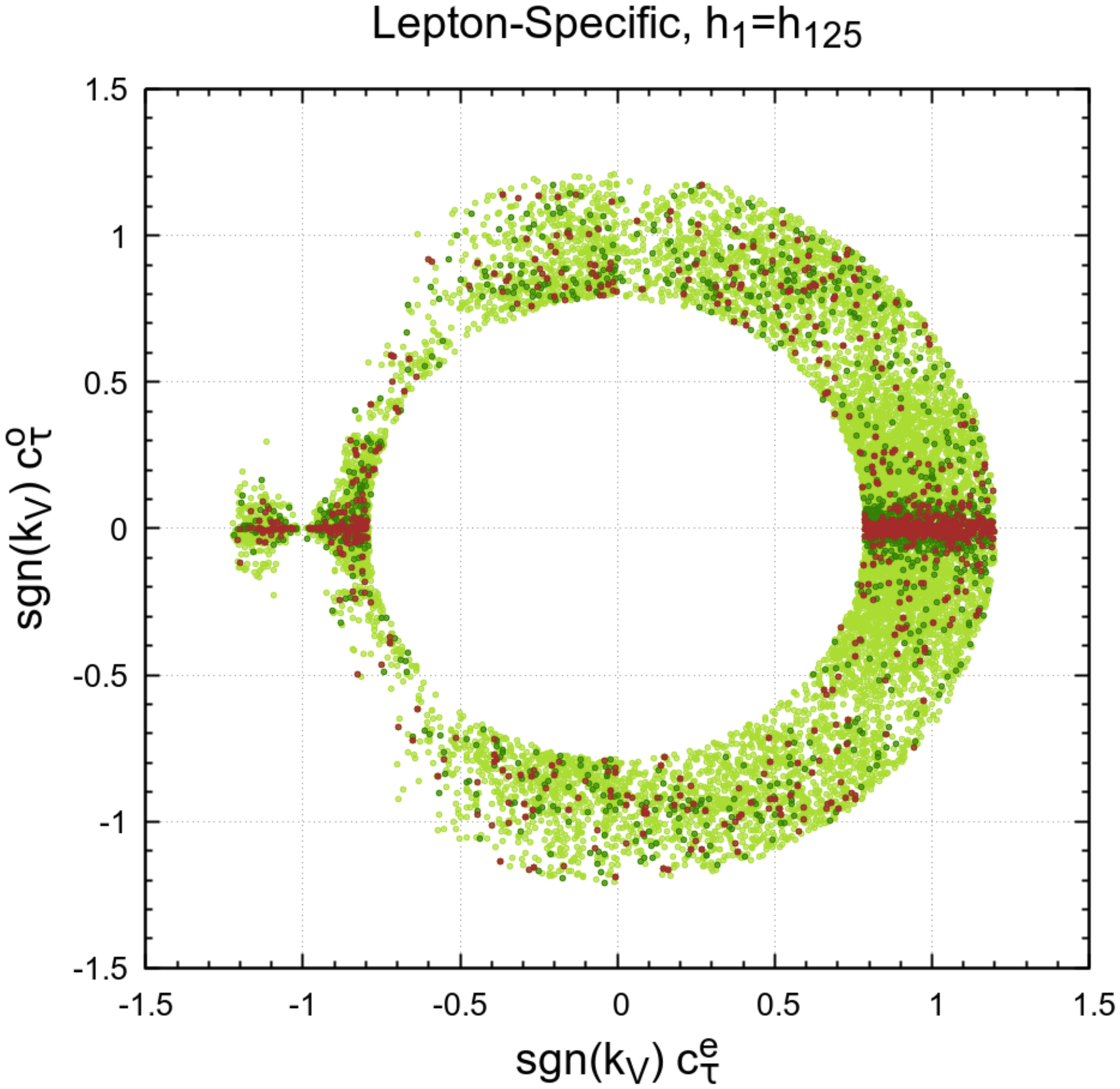}&
    \includegraphics[width=0.45\textwidth]{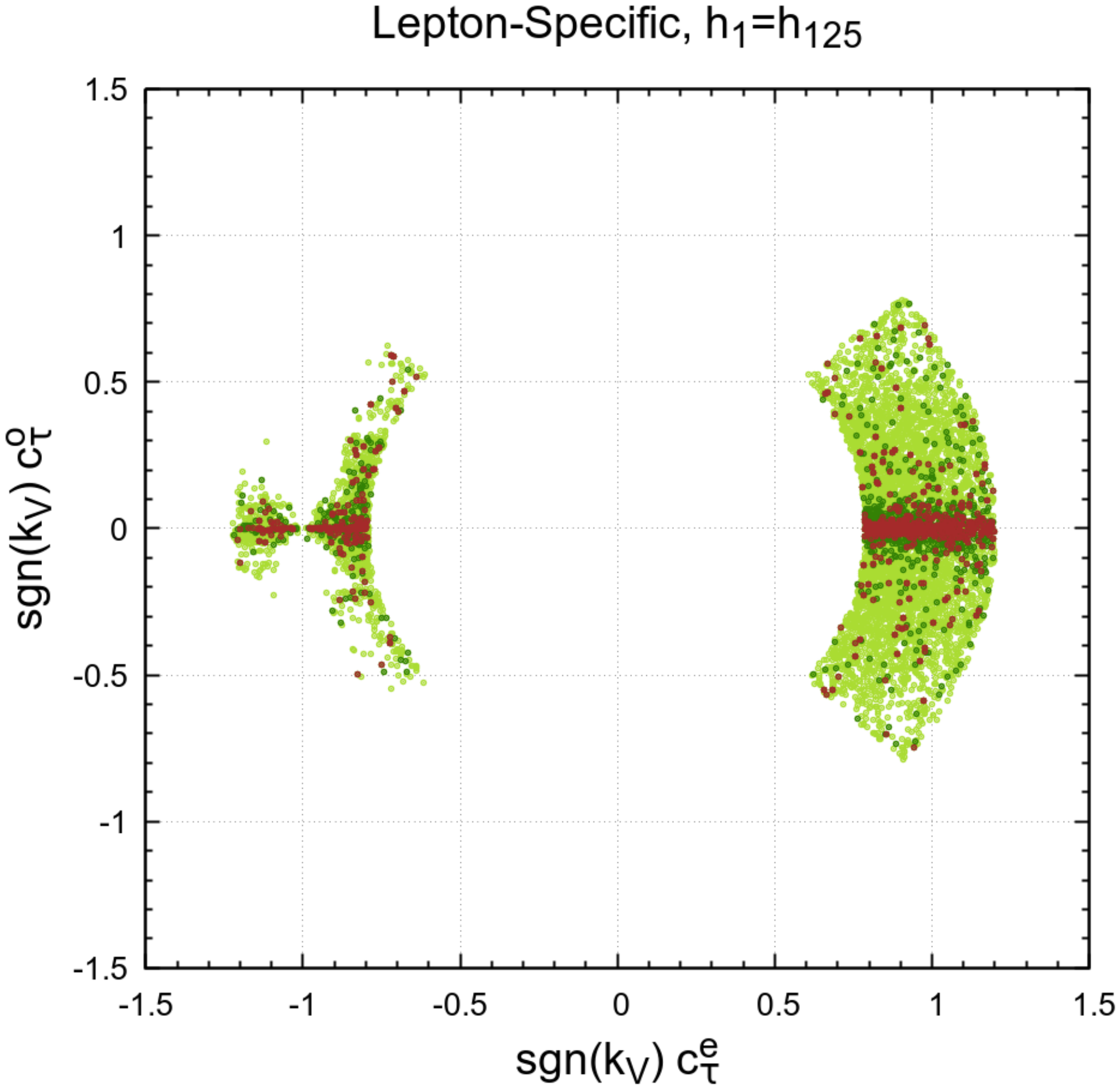}
  \end{tabular}
  \caption{CP-odd \textit{vs.}~CP-even component in the
    $h_{125}\tau\bar{\tau}$ coupling for the allowed
    parameter points in the LS
    model, assuming $h_1=h_{125}$,
    using 13~TeV LHC Higgs data on $h_{125}$ collected
    until 2022 and constraints from BSM
    scalar searches included in HT-1.1.3.
    In the left panel, the
            limit $\alpha_{h\tau\tau} < 41^\circ$
    from angular correlations of $\tau$ leptons
    in $h_{125} \to \tau \bar \tau$ decays
    is not applied, whereas the
    right panel includes this limit. Colour code as in Fig.~\ref{fig:1}.} 
  \label{fig:3}
\end{figure}

The results for the LS model with $h_1=h_{125}$ are shown in
\figref{fig:3}. This figures takes into account up-to-date constraints from LHC searches for additional Higgs bosons  and the signal rate measurements
of~$h_{125}$. Whereas in the left panel we exclude the constraints on the CP-violating phase $\alpha_{h\tau\tau}$ coming from direct searches for CP-violation in angular variables of $\tau$ leptons in $h_{125} \to \tau \bar \tau$ decays, reported by CMS~\cite{CMS:2021sdq}, in the right panel we include them. We see that, in the left panel, even though we are including the latest eEDM data, there is still a large allowed parameter region consistent with $|c^o_{\tau}| \gg |c^e_{\tau}|$.
On the other hand, the right panel shows that the constraints on $\alpha_{h\tau\tau}$ exclude this scenario.
Therefore, in the LS type of the C2HDM, the direct
LHC measurements of CP-violation in angular variables
in $h_{125} \to \tau \bar \tau$ decays are able to
exclude regions of the parameter space that would otherwise
be allowed by all other theoretical and experimental
constraints. In particular, we find parameter
points which are
in agreement with the experimental upper bounds
on the eEDM (dark red points in Fig.~\ref{fig:3}), and which are excluded only by the LHC
measurements of CP-violation
in decays of the
$h_{125}$ to $\tau$ leptons. This demonstrates
the complementarity of probing possible CP-violation
in extended Higgs-sector models at low-energies in
terms of eEDMs and at high-energy at the LHC.

Even though the direct limits on $\alpha_{h\tau\tau}$
substantially restrict the possible amount of
CP-violation in the $h_{125} \tau \bar \tau$ coupling,
they do not exclude the
interesting possibility $|c^o_{\tau}| \simeq |c^e_{\tau}|$, as can be seen in the right panel of \figref{fig:3}.
This has important consequences for the possibility
of explaining the baryon asymmetry of the Universe
in the framework of the C2HDM by means of
electroweak baryogenesis. It has recently been shown
that a sizable amount of CP-violation in the
$h_{125} \tau \bar \tau$ coupling might be sufficient
to accommodate the baryon asymmetry without
the presence of additional CP-violation in the
other couplings 
of the~$h_{125}$~\cite{Bahl:2022yrs}.\footnote{Compare, however,
also with \bib{Basler:2021kgq}.
Note furthermore that it is still under debate if the amount
of generated baryon asymmetry is sufficiently large to be in agreement with the observed value because
of large theory uncertainties in the prediction for the baryon asymmetry, which strongly depends on the
approach that is used to compute the source term for the baryon asymmetry,
see \bib{Postma:2022dbr} for a recent discussion.}
As such, the LS C2HDM can still be regarded as a
possible framework for an explanation of the
matter-antimatter asymmetry of the Universe.
We note, however, that a successful realization of
electroweak baryogenesis also requires a sufficiently
strong first-order electroweak phase transition.
We leave for future work an analysis of whether the allowed parameter points
in our scan sample that feature sizable CP-odd
components in the $h_{125} \tau \bar \tau$ coupling
can additionally accommodate such a phase transition.

\subsubsection{$h_2=h_{125}$}

\begin{figure}[htb]
  \centering
  \begin{tabular}{cc}
    \includegraphics[width=0.45\textwidth]{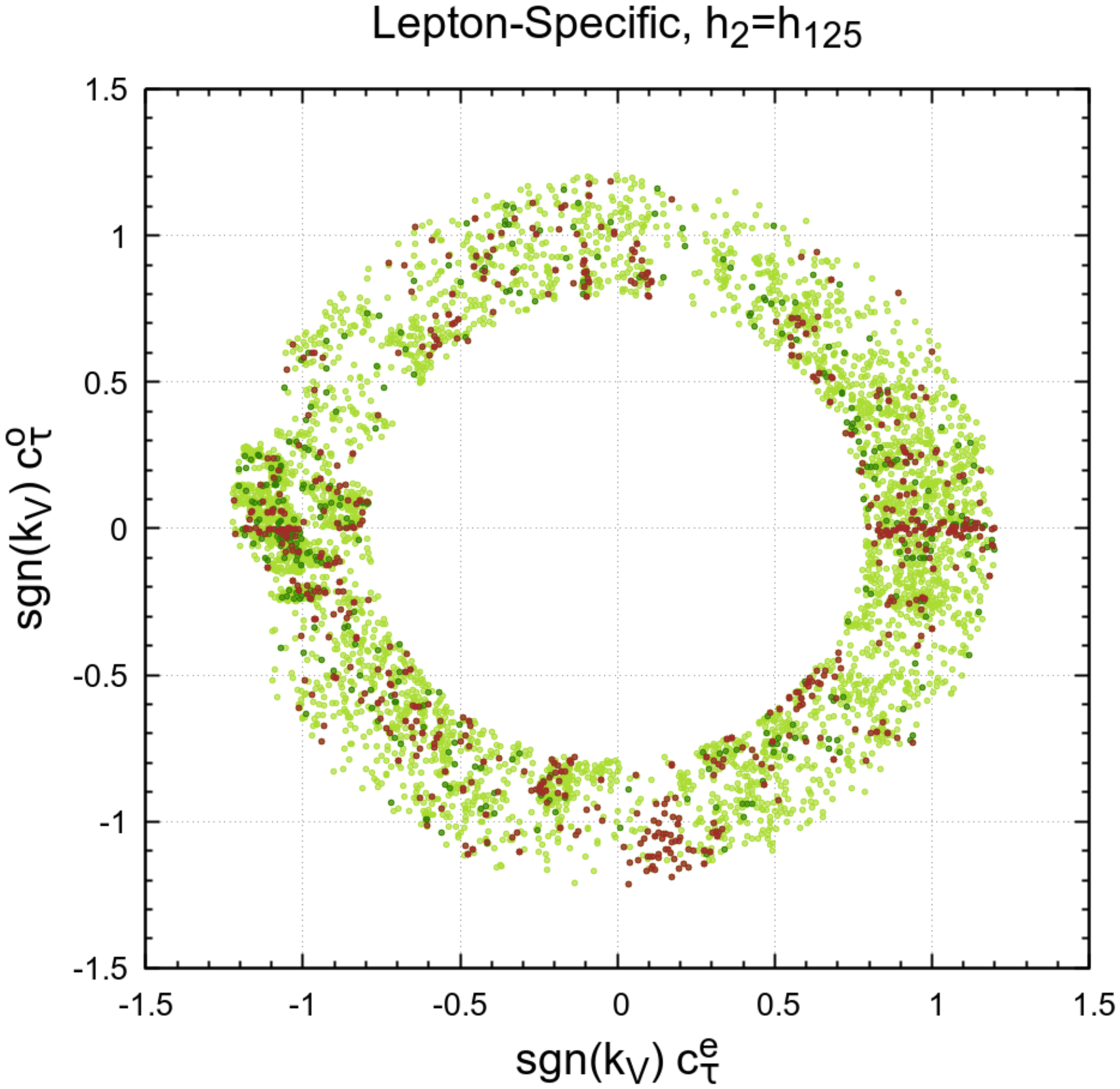}&
    \includegraphics[width=0.45\textwidth]{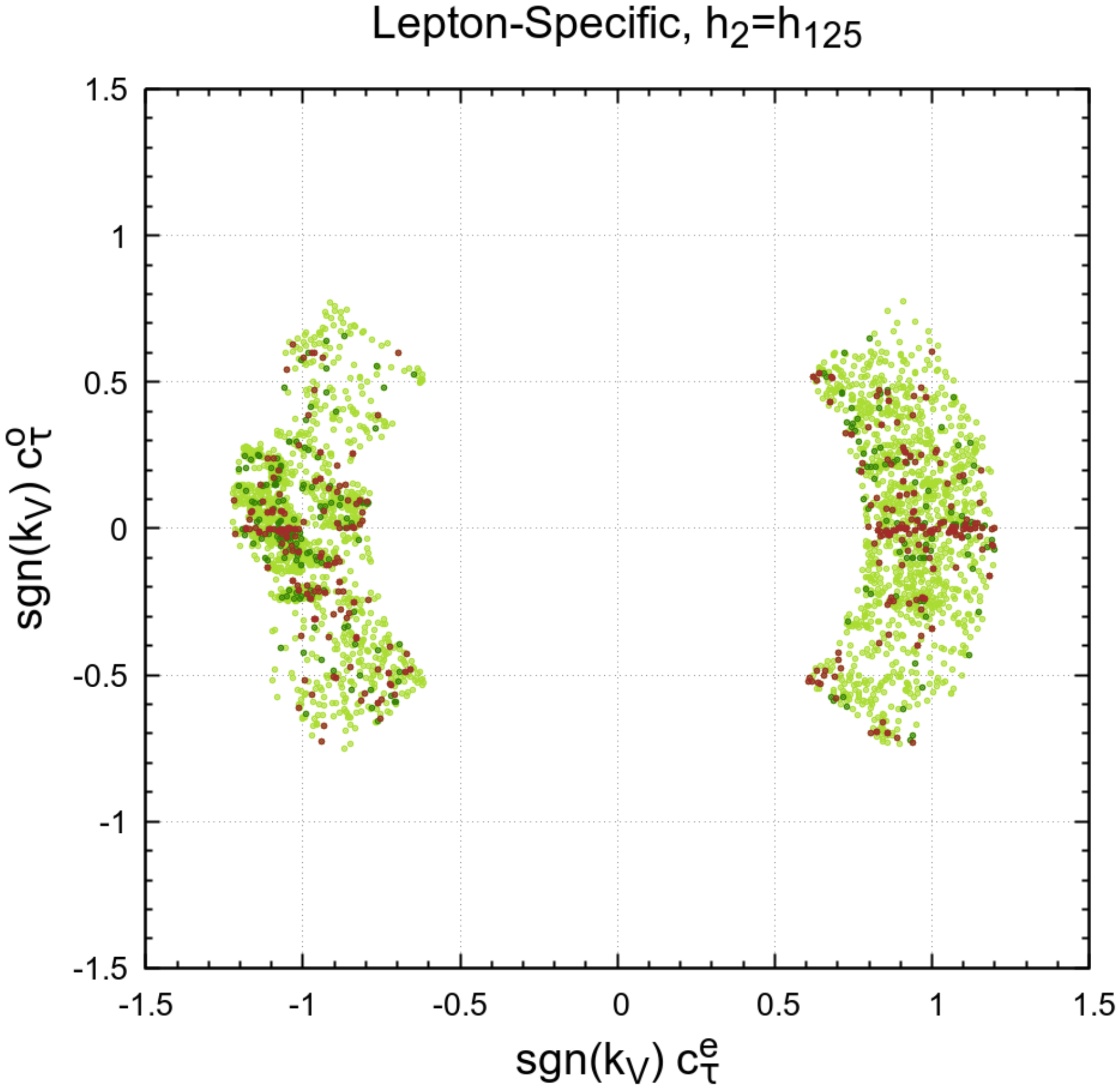}
  \end{tabular}
  \caption{Same as in \figref{fig:3}, but for $h_2=h_{125}$.}
  \label{fig:4}
\end{figure}

The case 
with the second lightest neutral scalar
$h_2$ acting as $h_{125}$ is similar to the case
where $h_1 = h_{125}$,
as can be seen in \figref{fig:4}. Here, we
show the same as in \figref{fig:3}, but now for
$h_2 = h_{125}$.
As in the previous case,
a parameter region with $|c^o_{\tau}| \gg |c^e_{\tau}|$ still remains
after the application of all current experimental
constraints from the LHC and the eEDM.
Consequently, also the mass hierarchy with
$h_2 = h_{125}$ is subject to new constraints
on the parameter space coming from the LHC
measurement of CP violating effects in
$h_{125} \to \tau \bar \tau$ decays, still leaving the interesting possibility of $|c^o_{\tau}| \simeq |c^e_{\tau}|$ though.

\subsubsection{$h_3=h_{125}$}

\begin{figure}[t]
  \centering
  \begin{tabular}{cc}
    \includegraphics[width=0.45\textwidth]{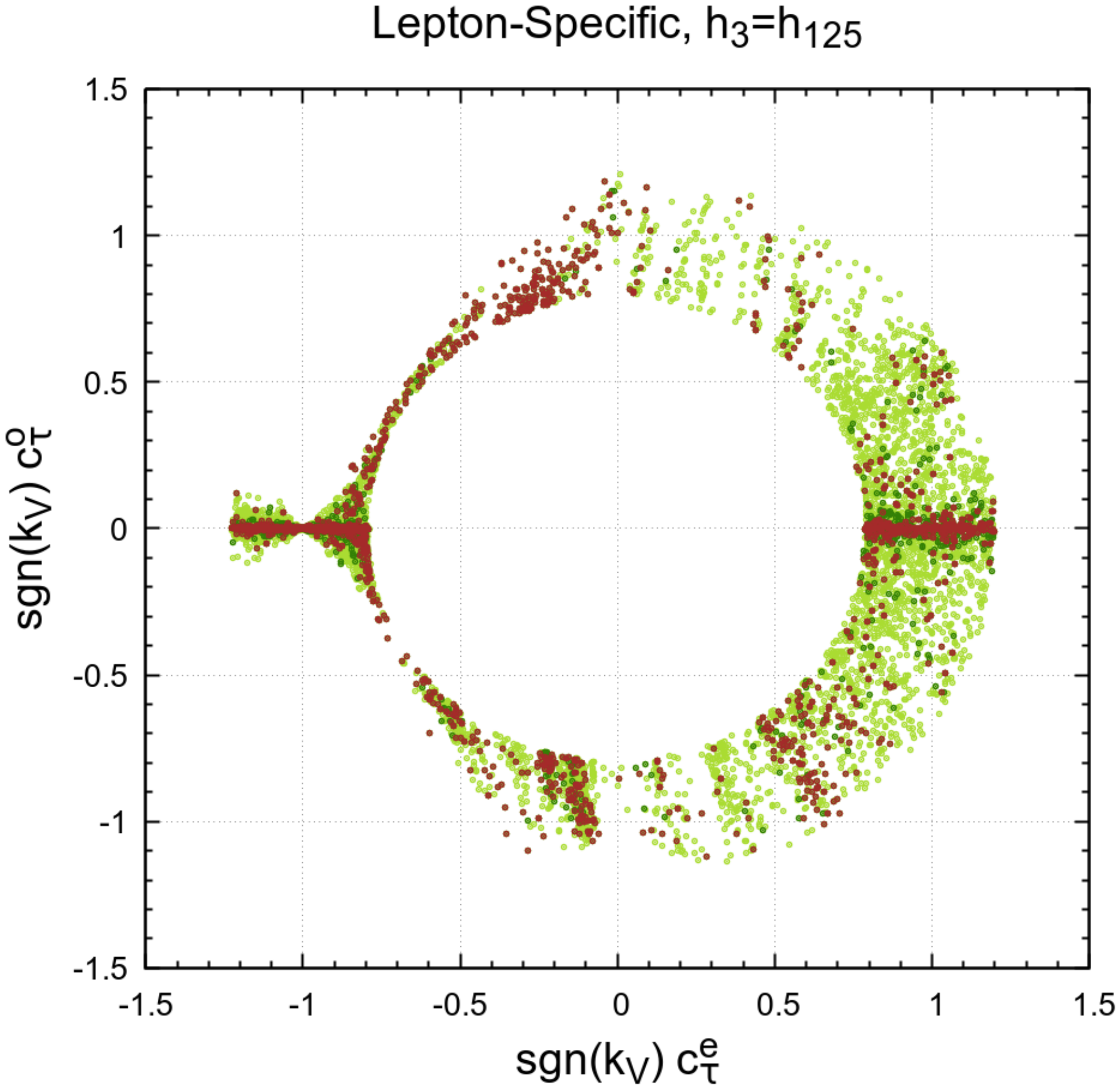}&
    \includegraphics[width=0.45\textwidth]{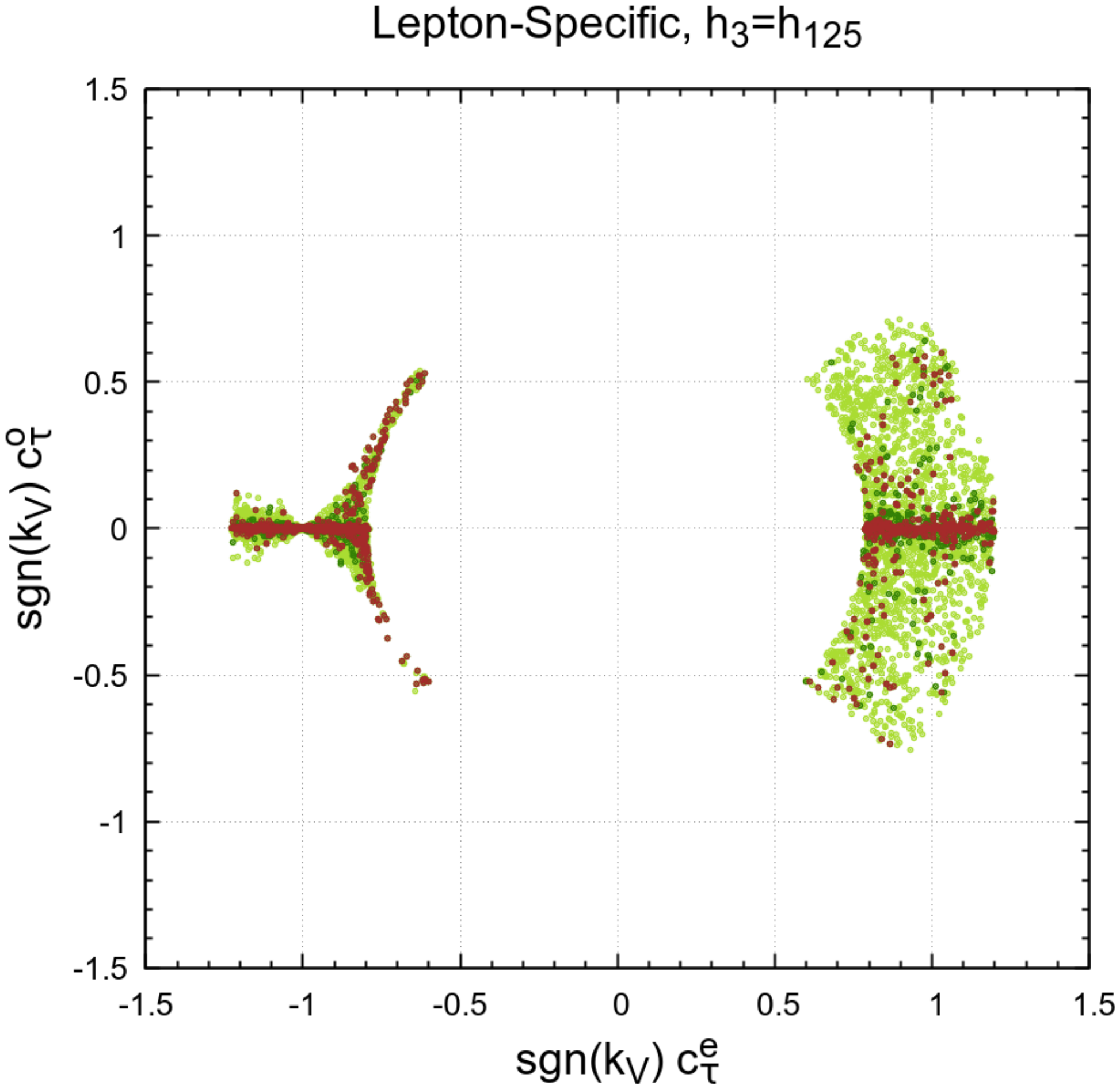}
  \end{tabular}
  \caption{Same as in \figref{fig:3}, but for $h_3=h_{125}$.}
  \label{fig:5}
\end{figure}

The situation for the 
mass hierarchy with $h_3 = h_{125}$,
is shown in \figref{fig:5}, where the
color of the points is defined as in Figs.~\ref{fig:3} and \ref{fig:4}.
Likewise, the limit on $\alpha_{\tau\tau}$
has been applied in the right plot only,
whereas all other constraints have been applied
in both plots according to the discussion above.
The main conclusions are the same as in the
previous two cases.
Since all mass hierarchies in the LS model
are compatible with CP-odd components
in the $h_{125} \tau \bar \tau$ coupling
that are sufficiently large to be directly observable
at the LHC, a possible future detection
of CP-violation in $h_{125} \to \tau \bar \tau$
decays via angular correlations
of the $\tau$ leptons would not
allow for a distinction between the different
mass hierarchies.

\subsection{\label{subsec:F}Flipped}

We have seen that the Higgs data related to $\mu_{VV}$ (as well as
direct searches~\cite{ATLAS:2020ior}) 
forbid values of $|c_t^0| \gtrsim |c_t^e|$,
and that the latest data of searches for CP-violation in angular
correlations of the 
$\tau$ leptons in $ h_{125} \rightarrow \tau \bar{\tau}$ decays
precludes $|c^o_{\tau}| \simeq
1$~\cite{CMS:2021sdq,ATLAS:2022akr}. 
 The Flipped type, with $\Phi_u = \Phi_\ell \neq \Phi_d$, might still be a promising candidate for large CP-odd components, as it
in principle allows
the possibility of large $|c^o_{b}|$. 
Moreover, since this type was shown in 2017
(after the 7~and 8~TeV Runs of the LHC) to
be able to accommodate sizable CP-odd components
with the mass hierarchy in which the detected Higgs boson
at 125~GeV is the lightest scalar~\cite{Fontes:2017zfn},
the impact of the more stringent
lower bounds on~$m_{H^\pm}$ from $b \to s \gamma$ transitions
can be expected to be less severe compared to the situation in Type-II
(see discussion in Sec.~\ref{subsec:Type-II}).

\begin{figure}[!h]
  \centering
  \begin{tabular}{cc}
    \includegraphics[width=0.45\textwidth]{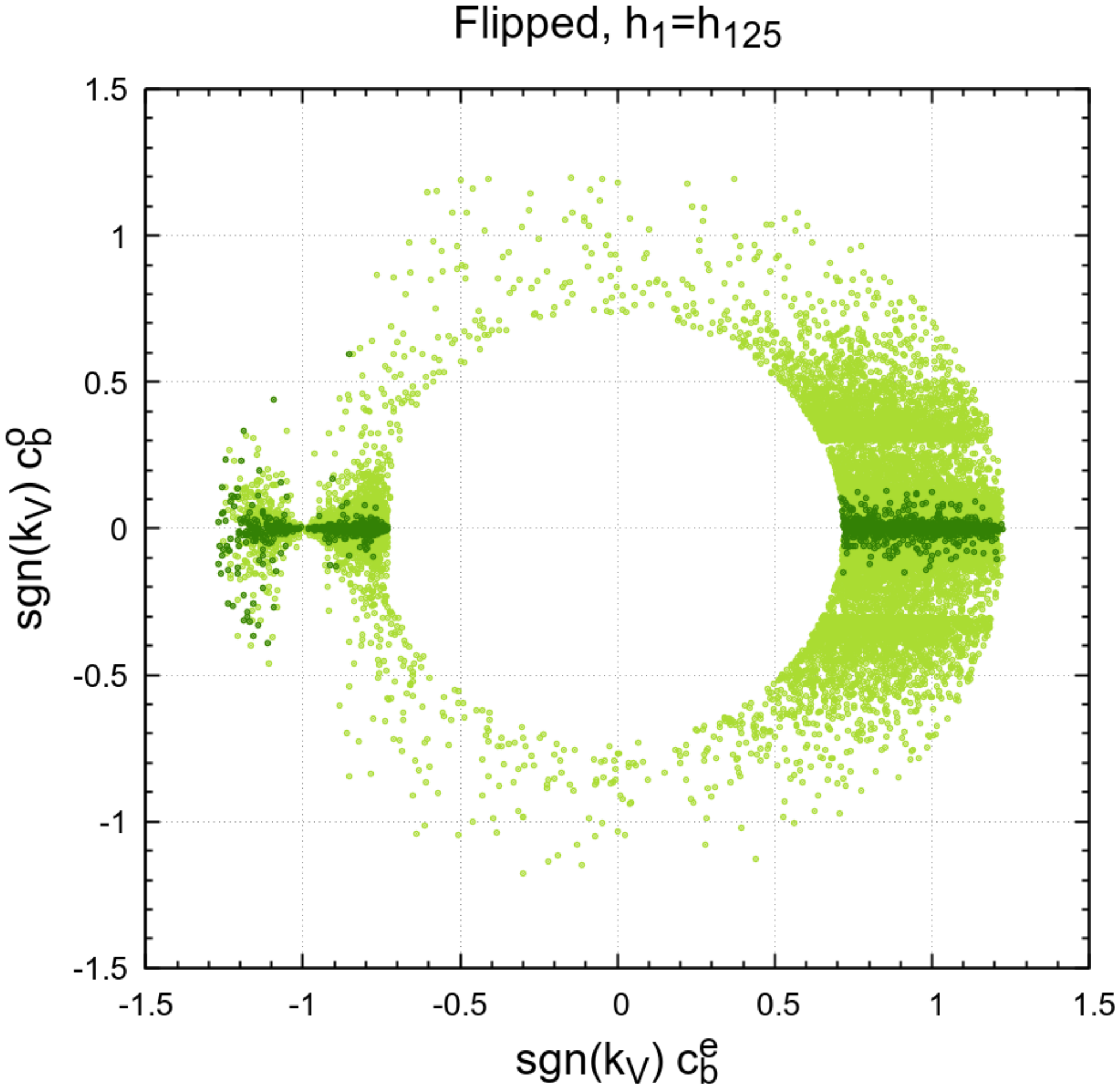}&
    \includegraphics[width=0.45\textwidth]{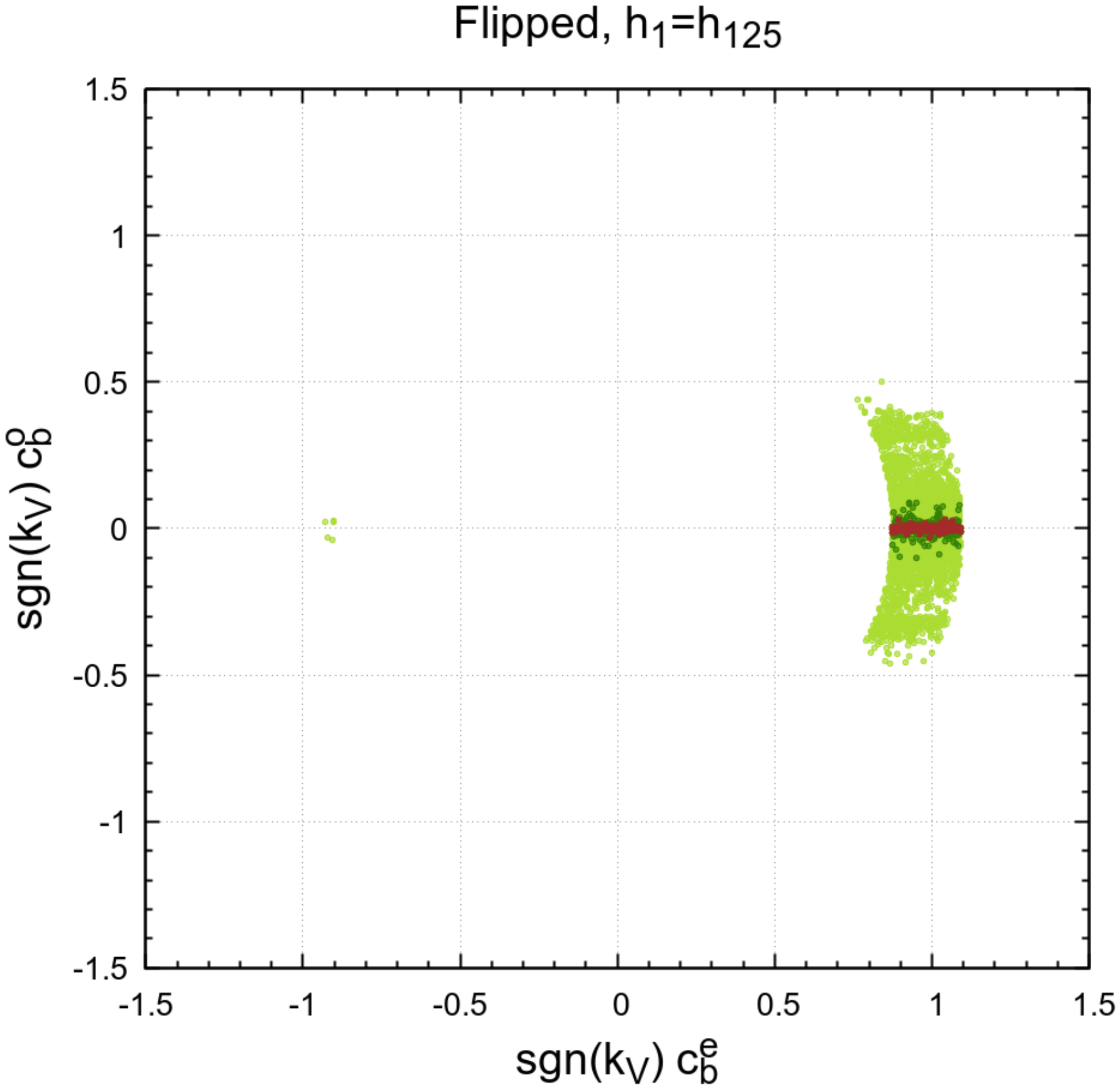}
  \end{tabular}
  \caption{CP-odd \textit{vs.}~CP-even component in the
    $h_{125}b\bar{b}$ coupling for allowed parameter
    points in the Flipped model, assuming
    $h_1=h_{125}$. Left panel: LHC 2017 data
    on $h_{125}$ and constraints from BSM
    scalar searches at 7~and 8~TeV included
    in HB-4.3.1. Right
    panel: LHC 2022 data on $h_{125}$,
    constraints from BSM scalar searches
    including searches at 13~TeV using
    HT-1.1.3 and the latest eEDM limit.
    Colour code
    as in Fig.~\ref{fig:1}.}
  \label{fig:7}
\end{figure}

It turns out that this possibility is now excluded. Indeed, the status
of the Flipped type changed significantly with respect to
\bib{Fontes:2017zfn}. This is not because of the LHC 2022 data on the
signal strengths of $h_{125}$, but because of the
additional constraints imposed by the
searches for extra scalars at 13~TeV,
as implemented in the most recent version
of HB (now incorporated in HT).  This is shown in \figref{fig:7},
which takes $h_1 = h_{125}$.
In the left panel, we use the old 2017 data on $h_{125}$ and the old
HB-4.3.1 bounds. 
In the right panel, we use HT-1.1.3 and the latest LHC 2022 data on searches for additional Higgs bosons, as well as the most recent limit on the eEDM~\cite{Roussy:2022cmp} (not used with the old data on the left panel).
The main impact of taking into account the 13~TeV signal-rate
measurements of~$h_{125}$ is that it decreases the width of the rings
on which allowed parameter points can be found.
We see a more 
significant difference 
between the left and the right plot of
\figref{fig:7} as a result of new constraints
from LHC searches for additional scalars at~13~TeV.
In particular, searches for one heavy Higgs boson decaying into a $Z$ and
another Higgs boson, both by ATLAS~\cite{ATLAS:2018oht} and
CMS~\cite{,CMS:2019qcx}, together with the latest eEDM results --- precludes the situation where 
$c^e_{b} \simeq 0$,
which is thus not visible in the right panel
of \figref{fig:7}. 
One should note here that, in the CP-conserving limit
of the 2HDM with $h_{125}$ predicted to
be CP-even as in the SM,
the decay $h_i \to Z h_{125}$ is only
allowed for a CP-odd state $h_i$, whereas
there is no coupling between a $Z$ boson and
two CP-even scalars. If, on the other hand,
$h_{125}$ carries a CP-odd admixture,
both heavier neutral scalars $h_2$ and $h_3$
can decay into a
$Z$ boson and $h_{125}$. It follows that searches for heavier Higgs bosons decaying
into the 125~GeV Higgs boson and a $Z$-boson are
exceptionally important if CP-violation is
present in the scalar sector
(see also Refs.~\cite{Fontes:2015xva,Chen:2015gaa,
Chen:2017com}). In our case, indeed, they exclude large parts of \figref{fig:7} with $|c_b^e|<1$. This is a confirmation of the important physical insight gained during Run~2 and, in this particular instance, on the
crucial new bounds placed on the production
of additional Higgs bosons.

As a result of the application of the
13~TeV BSM scalar searches,
there is no further impact from the CMS constraints on $\alpha_{h\tau\tau}$~\cite{CMS:2021sdq}.
We observe that, at the current level of experimental
precision, the direct limit
on $\alpha_{h\tau\tau}$
does not yet play
a role in the CP properties of the coupling $h_{125}b\bar{b}$. This is expected by the fact
that, in the Flipped type, the $h_{125}\tau\bar\tau$
and the $h_{125}b\bar b$ couplings are
independent parameters, according to
$\Phi_u = \Phi_\ell \neq \Phi_d$.
In the end, then, the situation that was
shown to be possible in \bib{Fontes:2017zfn} (left panel of
\figref{fig:7}) is reduced to almost vanishing 
CP-odd components, $c_b^o$. These would not be
observable directly at the LHC, 
due to the combination of the
new results from the eEDM (dark red points)
and searches for additional
scalars at the LHC.

\section{\label{sec:conclusions}Conclusions}

The C2HDM is one of the simplest extensions of the SM with one new
CP-violating parameter, which arises from the scalar sector.  The
experimental data available in 2017,
in particular the LHC data collected at
7~and 8~TeV, still allowed for the striking
possibility that the CP-odd components ($c^o$) of the $h_{125}$
couplings to down-type quarks and/or charged leptons were much larger
than the corresponding CP-even components ($c^e$). This was possible
for all types, except Type-I~\cite{Fontes:2017zfn}.
The possible presence of sizable CP-odd coupling components would be a
clear indication of physics beyond the~SM, and it could give rise to
important phenomenological consequences.  For instance, the viability
of a realization of electroweak baryogenesis in generic extensions of
the SM by a second Higgs doublet field relies on additional sources of
CP-violation according to the Sakharov 
conditions~\cite{Sakharov:1967dj}.

In this paper, we re-analyze the situation in light of the most recent
experimental constraints, namely: up-to-date results from LHC's Run 2
at~13~TeV regarding searches for additional scalars and the
signal-rate measurements of the Higgs boson at 125~GeV, new eEDM
results, data from direct searches for CP-violation in angular
correlations of final state $\tau$-leptons in $h_{125} \rightarrow
\tau \bar{\tau}$ decays in terms of the effective mixing angle
$\alpha_{h\tau\tau}$ and, finally, discussing the bounds on
$m_{H^\pm}$ coming from the $b\to s \gamma$ constraints.
The current situation is summarized in \tab{tab:3}.
\begin{table}[h]
  \centering
  \begin{tabular}{|l|c|c|c|c|}\hline
    Type   & I & II& LS& Flipped\\ \hline
$h_1=h_{125}$&$\times$ &$\times$ &$\tau$ & $\underline{\times}$\\\hline
$h_2=h_{125}$ &$\times$ & $\underline{\times}$ &$\tau$ & $\times$\\\hline
$h_3=h_{125}$ &$\times$ &$\times$ &$\tau$ & $\times$\\\hline
  \end{tabular}
  \caption{Current results for the large Yukawa couplings.
A cross means that it is not possible to have
large CP-odd couplings, i.e.~$|c^0| \gtrsim |c^e|$.
The notation $\tau$ means that $c^o/c^e$ is
limited by the direct
searches for CP-violating angular correlations
of $\tau$ leptons in
$ h_{125} \rightarrow \tau \bar{\tau}$
decays~\cite{CMS:2021sdq}.
Underlined crosses indicate a change
from allowed ($\checkmark$) to excluded ($\times$)
compared
to the previous analysis carried out
in 2017~\cite{Fontes:2017zfn}.
}
  \label{tab:3}
\end{table}
This can be compared with \tab{tab:2} summarizing the status in 2017
found in a previous analysis~\cite{Fontes:2017zfn}.  The most
important conclusions of the updated analysis presented here are the
following:
\begin{description}
\item[Type-II:]
Previously, the possibility of sizable CP-odd
components in the couplings of~$h_{125}$,
$|c_b^o|\gg |c_b^e|$, remained experimentally
viable for the mass hierarchy with $h_{125}$ being the second lightest
neutral scalar ($h_2=h_{125}$)~\cite{Fontes:2017zfn}.
This possibility is now practically
excluded, as we discussed
in Sec.~\ref{subsec:Type-II}.  This exclusion follows from the combination of
stringent eEDM constraints and 
limits from LHC searches for additional Higgs bosons
at 13~TeV, most notably  from searches for
di-$\tau$ resonances below 125~GeV~\cite{CMS:2022goy}.
The only still viable option for sizable $|c_b^o|$ happens if two conditions are verified: first, if the parameters are very fine-tuned, such that
cancellations between different contributions to the eEDM occur at (or below) the percent-level; second, if the pole masses for top-quark and $\tau$-lepton
and the running bottom-quark mass $\overline{m}_b(\overline{m}_b)$ are 
used for the computation of the eEDM.
If both conditions are met, the LHC measurements of $\alpha_{h\tau\tau}$
exclude otherwise allowed parameter space.
If, however, the running fermion masses at the scale $M_Z$
are used for the computation of the eEDM,
no parameter points can be found that
simultaneously evade the experimental upper bound on the eEDM, and comply with the LHC cross-section limits from BSM scalar searches. This strong dependence 
on the fermion masses
emphasizes that, in order to conclude whether Type-II can accomodate sizable CP-violating Higgs-boson couplings, one needs a proper understanding of which
prescription to use for the fermion masses
in the calculation of the eEDM.
\item[LS:] This Yukawa type
is the only one that can still accommodate
sizable CP-odd components in the couplings
of~$h_{125}$
to charged leptons, while being in agreement
with all theoretical and experimental
constraints. In particular,
the possibility of $|c_\tau^o| \simeq |c_\tau^e|$ is still
allowed in the LS case, and only in this case.
We find values of
$\alpha_{h\tau\tau}$ that would be directly
observable at the LHC by measurements of
angular correlations of final state $\tau$
leptons in $h_{125} \to \tau\bar\tau$ decays.
Consequently, in the LS type,
the recently reported $2\sigma$ confidence-level limits
of $\alpha_{h\tau\tau} < 41^\circ$
from CMS~\cite{CMS:2021sdq}
and of $\alpha_{h\tau\tau} < 34^\circ$
from ATLAS~\cite{ATLAS:2022akr}
give rise to new constraints on the C2HDM parameter space, excluding
previously allowed parameter space regions.
According to our findings, if in the future a non-vanishing value of
$\alpha_{h\tau\tau}$ were measured at
the LHC, this would point
towards the LS~type, allowing to experimentally distinguish this type
from the other Yukawa types of the C2HDM.
Since all possible mass hierarchies of neutral scalars
were shown to be compatible with sizable values of
$\alpha_{h\tau\tau}$, a possible future detection
of a CP-violating $h_{125} \tau \bar \tau$
coupling would not decide whether the detected
Higgs boson at 125~GeV would correspond to
the lightest, the second-lightest
or the heaviest neutral scalar of the C2HDM.
\item[Flipped:] In this type, one has
$c_\tau^o = c_t^o$. Hence, the circumstance that $|c_t^o|$ is already 
stringently constrained from signal-rate measurements
of~$h_{125}$ renders the aforementioned constraints on $\alpha_{h\tau\tau}$ irrelevant.
On the other hand, the possibility $|c^o_b| >
|c^e_b|$ was previously allowed, assuming that
the 125~GeV Higgs boson is the lightest neutral scalar. Here, we
demonstrated that this possibility is now also forbidden in this type
of the C2HDM, due to the LHC's improved bounds from searches for extra
scalars (in combination with the other experimental constraints). The
most relevant searches 
are those involving 
one heavy Higgs boson decaying into a $Z$ and
another Higgs boson, both by ATLAS~\cite{ATLAS:2018oht} and
CMS~\cite{,CMS:2019qcx}. Additionally, the more recent eEDM
bounds~\cite{Roussy:2022cmp} constrain
$c^o_b$ to lie very close to zero.
\end{description}

In summary, we have shown that the possibility of 
sizable CP-odd components
${|c^o| \simeq |c^e|}$ in the couplings of the
125~GeV Higgs boson is only
allowed in the LS cases (all mass orderings),
where the CP-violation appears in the
couplings of the Higgs boson to $\tau$ leptons.
The possible amount of CP-violation
is then limited ultimately
by the direct searches for CP-violation in
angular correlations between $\tau$ leptons
produced in Higgs boson decays.
These measurements have been performed by
both ATLAS and CMS utilizing the full Run~2
dataset. The measurements are currently statistically
limited.
The anticipated
future improvements on their experimental precision
will be paramount to our understanding of the C2HDM and
its phenomenology at the LHC, as well as of the extent
to which the shortcomings of the SM can be addressed
in this model.

\section*{Acknowledgments}
T.B.~thanks Henning Bahl for useful discussions.
The work of J.C.R.~and J.P.S.~is supported in part
by the Portuguese Funda\c{c}\~{a}o
para a Ci\^{e}ncia e Tecnologia\/ (FCT) under Contracts
CERN/FIS-PAR/0002/2021, UIDB/00777/2020, and UIDP/00777/2020\,;
these projects are partially funded through POCTI (FEDER),
COMPETE, QREN, and the EU.
D.F.~is supported by the U.S.~Department of Energy under Grant
Contract No.~DE-SC0012704.
R.S.\ is partially supported by FCT under Contracts no.\ UIDB/00618/2020, UIDP/00618/2020, CERN/FIS-PAR/0025/2021, CERN/FIS-PAR/0021/2021 and CERN/FIS-PAR/0037/2021.
M.M.~acknowledges financial support from the Deutsche Forschungsgemeinschaft (DFG, German Research Foundation) under grant 396021762 - TRR 257.
The work of T.B.~is supported by the German
Bundesministerium f\"ur Bildung und Forschung (BMBF, Federal
Ministry of Education and Research) -- project 05H21VKCCA.

\bibliographystyle{JHEP}
\bibliography{paper}

\end{document}